\documentclass[journal]{IEEEtran}
\usepackage{graphicx}
\usepackage{multirow}
\usepackage{hyperref}
\usepackage{cite}
\usepackage{lineno}
\usepackage{amsmath,amsfonts,amssymb}
\usepackage{subfig}
\usepackage{makecell}
\usepackage[noend]{algpseudocode}
\usepackage{algorithmicx,algorithm}
\usepackage[numbers]{natbib}

\begin{document}
\title{Seismic Traveltime Tomography with Label-free Learning}
\author{Feng~Wang, Bo~Yang, Renfang~Wang and Hong~Qiu
\thanks{This work was supported in part by the National Natural Science Foundation of China under Grant 41774079 and Grant 6190617, in part by the Project of the Science and Technology Plan for Zhejiang Province under Grant LGF21F020023 and ZCLY24F0301, and in part by the Plan Project of Ningbo Municipal Science and Technology under Grant 2021Z050, Grant 2022Z233, Grant 2022S002, and Grant 2023J403. (\emph{Corresponding author: Bo~Yang})}
\thanks{F.~Wang, R.~Wang and H.~Qiu are with the College of Big Data and Software Engineering, Zhejiang Wanli University, Ningbo 315100, China (e-mail: wangf\_721@zju.edu.cn; renfang\_wangac@126.com; qiuhong@zwu.edu.cn).}
\thanks{B.~Yang is with the Key Laboratory of Geoscience Big Data and Deep Resource of Zhejiang Province, School of Earth Sciences, Zhejiang University, Hangzhou310027, China (e-mail: bo.yang@zju.edu.cn).}}
\maketitle

% \linenumbers

\begin{abstract}
Deep learning techniques have been used to build velocity models (VMs) for seismic traveltime tomography and have shown encouraging performance in recent years. However, they need to generate labeled samples (i.e., pairs of input and label) to train the deep neural network (NN) with end-to-end learning, and the real labels for field data inversion are usually missing or very expensive. Some traditional tomographic methods can be implemented quickly, but their effectiveness is often limited by prior assumptions. To avoid generating and/or collecting labeled samples, we propose a novel method by integrating deep learning and dictionary learning to enhance the VMs with low resolution by using the traditional tomography-least square method (LSQR). We first design a type of shallow and simple NN to reduce computational cost followed by proposing a two-step strategy to enhance the VMs with low resolution: (1) Warming up. An initial dictionary is trained from the estimation by LSQR through dictionary learning method; (2) Dictionary optimization. The initial dictionary obtained in the warming-up step will be optimized by the NN, and then it will be used to reconstruct high-resolution VMs with the reference slowness and the estimation by LSQR. Furthermore, we design a loss function to minimize traveltime misfit to ensure that NN training is label-free, and the optimized dictionary can be obtained after each epoch of NN training. We demonstrate the effectiveness of the proposed method through the numerical tests on both synthetic and field data. 
\end{abstract}

\IEEEpeerreviewmaketitle

\section{Introduction}
\IEEEPARstart{S}{eismic} traveltime tomography has been widely used to build VMs from the traveltimes between pairs of source and receiver to image the subsurface structure. It has been successfully applied to build VMs at different scales including local scale\cite{mordret_ambient_2014}, regional scale\cite{gorbatov_signature_2000} and global scale\cite{meier_global_2007}, and has also been used to produce images in near-surface exploration\cite{allmark_seismic_2018}. 

Tomography is generally regarded as an non-linear ill-posed inverse problem. Researchers have proposed two kinds of methods to solve this problem, including linearization and nonlinear inversion approaches. To find the solutions with minimal misfit, linearization inversion approaches require to linearized the tomography operator to simplify the inverse problem, such as LSQR, sparsity constrained inversion methods\cite{loris_nonlinear_2010} and dictionary learning\cite{bianco_travel_2018}. However, the linearization may produce large difference between linearized and true probabilistic solutions\cite{galetti_uncertainty_2015}. Although nonlinear inversion methods such as Monte Carlo can solve the inverse problems without linearization\cite{piana_agostinetti_local_2015,zhao_bayesian_2022}, the computational cost is significantly expensive. 

Deep learning utilizes deep NNs to learn the complex relationships and address the nonlinear ill-posed inverse problems by developing high-level representations of data using stacked layers of neurons and multiple nonlinear transformations\cite{mousavi_deep-learning_2022}, which makes deep learning a powerful numerical tool for solving the high dimensional nonlinear ill-posed problems. Therefore, deep learning has also become popular in the community of VMs building. Currently, the deep-learning-based velocity inversion methods can be broadly categorized as data-driven deep learning inversion and model-driven deep learning inversion. 

\emph{Data-driven deep learning inversion.} For data-driven deep learning inversion, it is usually necessary to first establish the training dataset (i.e., labeled samples), and then train the deep NNs in end-to-end manner. Forward simulation is currently the main means of obtaining training dataset as the real labeled samples is usually missing or very expensive. Once the training dataset is prepared, some classical NNs, such as fully connected network (FCN)\cite{moya_inversion_2010}, convolution neural network (CNN)\cite{araya-polo_deep-learning_2018}, U-net\cite{geng_deep_2022} and recurrent neural network (RNN)\cite{fabien-ouellet_seismic_2020}, can be trained to predict the VMs from observations (e.g., shot gathers). Generative adversarial network (GAN) \cite{goodfellow_deep_2016} is a kind of unsupervised learning methods that can utilize the unlabeled data in the training process. For example, \cite{cai_semisupervised_2022} developed a semi-supervised surface wave tomography with wasserstein cycle‐consistent GAN that takes both model-generated and observed surface wave dispersion data in the training process. Contrary to the methods that only provide deterministic solutions for inverse problems, deep-learning-based probabilistic inversion approaches can obtain the posterior probabilistic density function (pdf), which can be used to constitute the full solutions of inverse problem. \cite{devilee_efficient_1999} used NNs to provide posterior pdfs for discrete Bayesian tomography. \cite{earp_probabilistic_2020} introduced mixture density network into 2 dimensional (2-D)  traveltime tomography.

\emph{Model-driven deep learning inversion.} To alleviate the dependence of NN training on the amount of training dataset, model-driven deep learning inversion approaches have been developed, and shown encouraging performance. These types of approaches embed physical information into deep learning models, making the training can be implemented through a small amount of boundary conditions. As the representative model in the model-driven deep learning inversion, physic-informed neural network (PINN) \cite{raissi_physics-informed_2019} integrates the governing physics law into the learning process, and it has been widely used for solving partial differential equation (PDE), such as seismic wavefield modeling and traveltime tomography. \cite{song_solving_2021} solved the frequency-domain acoustic VTI wave equation using PINN. \cite{song_simulating_2022} proposed a Fourier feature PINN to overcome the problem of spectral bias for simulating multifrequency waveﬁelds. The eikonal equation plays an important role in traveltime tomography. \cite{waheed_pinneik_2021} proposed a PINN-based solver for solving the 2-D eikonal equation. \cite{taufik_upwind_2022} demonstrated that PINN can produce more accurate results than conventional approaches. To mitigate the uncertainty effects and quantify their impacts in the prediction, \cite{gou_bayesian_2022} proposed Bayesian PINN to infer the velocity field and reconstruct the traveltime field. \cite{grubas_neural_2023} solved the isotropic eikonal equation by improving accuracy of PINN. \cite{chen_eikonal_2022} presented an PINN-based eikonal tomography approach for Rayleigh wave phase velocities and applied it to regional scale.      

The above-mentioned deep-learning-based tomographic methods have shown the ability to outperform traditional approaches, but their performance still depend on 1) \emph{labeled samples or labels} and 2) \emph{large models}. In NNs training process, the NNs' parameters are optimized to minimize the misfit between the prediction and the corresponding label. Generally, the larger the training dataset, the better the NNs' generalization. However, the labels are non-existent or high expensive for real data inversion, and the synthetic training dataset generated by forward methods can not fully represent the distribution of the real data. Usually, large NN models can outperform the small ones that is the main reason why most of current deep-learning-based traveltime inversion methods tend to take large models (e.g., U-Net, LSTM, GAN and Transformer) as their backbone network, but large model training is computationally costly. In addition, although the end-to-end tomography can infer rapidly (i.e., predict VMs from observation directly), it ignores the underlying physical laws, which makes the predictions suffer from the black-box nature of NNs. 

It worth to point out that the deep dictionary learning (DDL)\cite{tariyal_deep_2016} is a novel framework which utilizes the advantages of both dictionary learning and deep learning to learn hierarchical features from data. Different from conventional deep NNs, DDL substitutes the "weights" or "filters" in NN with the "basis" and "features" by matrix factorization. This framework has been mainly applied to image classification and cluster, and it has achieved higher accuracy than conventional deep NN such as stacked autoencoder, deep belief network, and CNN\cite{tariyal_deep_2016,mahdizadehaghdam_deep_2019,tang_when_2021}. Unfortunately, the labels are still indispensable in the model training of DDL.      

 Traditional traveltime tomographic methods such as LSQR that can be implemented rapidly and do not require labeled samples, but their effectiveness depend on the assumptions or prior information. On the contrary, the deep-learning-based tomography can be independent on prior information. Therefore, this paper intends to integrate traditional methods and deep learning to obtain the high-resolution VMs without labels. We design a type of shallow and simple NN to reduce the computational cost. Instead of using end-to-end learning to predict the VMs directly from the observed traveltime, we use the NN to optimize the initial dictionary that learned from the estimated VMs by LSQR. We reconstruct the high-resolution VMs through the optimized dictionary, the estimated velocity by LSQR, and the reference slowness that serves as an initial guess for LSQR tomography. We train the NN by using the initial dictionary and observed traveltime. The objective is to minimize the traveltime MSE (mean square error) loss to avoid the requirement for the labeled samples. The NN and the initial dictionary are optimized simultaneously, which means that the NN can provide the optimized dictionary after each epoch of NN training. 

The organization of this article is as follows. We first set up the optimization problem for slowness perturbations, and then we propose a novel scheme to obtain high-resolution VMs using the combination of NN and dictionary learning followed by the NN designing and training. After that, we demonstrate the effectiveness of the proposed method by the numerical tests. Finally, we provide a brief discussion about the uniqueness of this work and draw conclusions. Source code is available at \url{https://github.com/linfengyu77/STTwLL}. 

\begin{figure}[]
\centering
\includegraphics[width=2.5in]{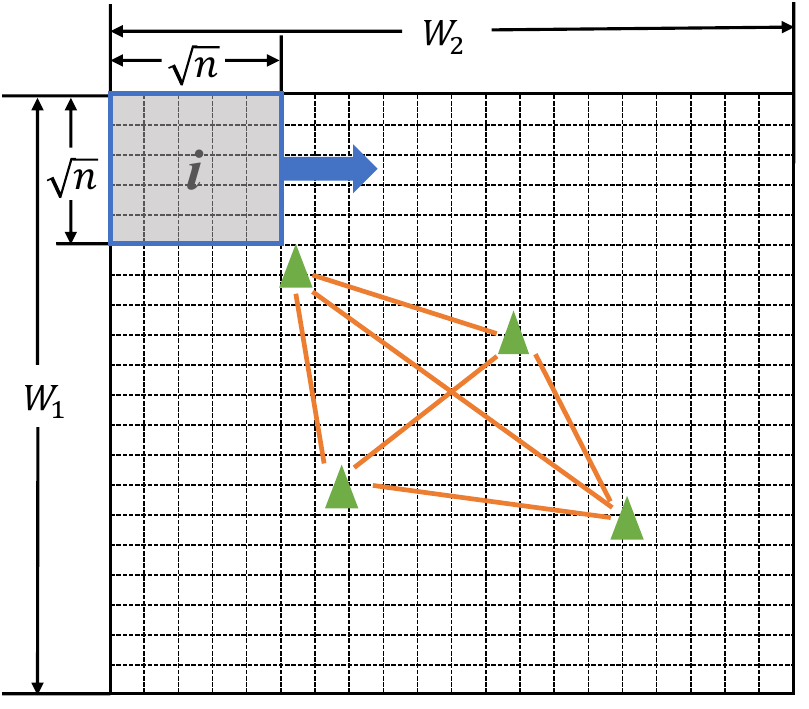}
\caption{2-D slowness image divided into pixels (dashed boxes) according to \cite{bianco_travel_2018}. $W_{1}$ and $W_{2}$ represent the number of pixels in vertical and horizontal directions, respectively. The green triangles are receivers and the orange lines represent the rays between them. The gray region represents the \emph{i-th} patch containing $n$ pixels.} 
\label{fig1}
\end{figure}

\begin{figure}[!tb]
\centering
\includegraphics[width=3in]{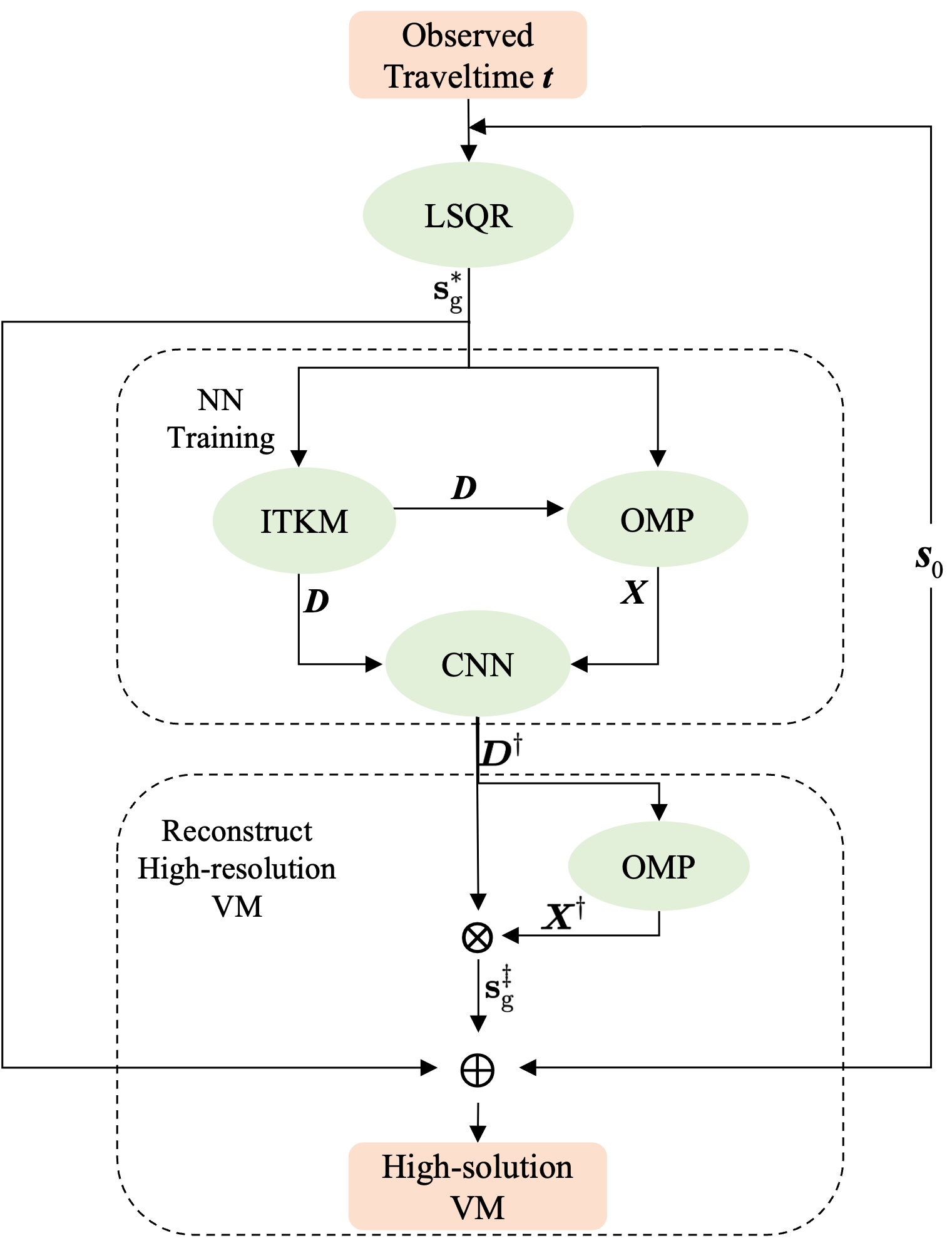}
\caption{Schematic diagram of our method. The symbol $\bigotimes$ is matrix multiplication. The vectors vector $\mathbf{t}$ and $\mathbf{s}_{\mathrm{g}}$ denote the traveltime and desired perturbations, respectively. $\mathbf{s}^{\ast}_{\mathrm{g}}$ is the perturbations inverted by LSQR, $\mathbf{D}$ is the initial dictionary trained by dictionary learning,  while $\mathbf{X}$ is the initial code obtained by sparse coding. $\mathbf{D}^{\dagger}$ is the dictionary optimized by NN, and $\mathbf{X}^{\dagger}$ is the code for $\mathbf{D}^{\dagger}$ by sparse coding.}
\label{method}
\end{figure}

\section{Methodology}
In this section, we present our approach. We first set up the MAP problem for improving the quality of the VMs with low resolution, and we then propose a two-step strategy to solve the MAP problem, which integrates dictionary learning and deep learning without training dataset.

\subsection{Problem setup}
In this paper, we consider the case of 2-D traveltime tomography for surface waves using seismic interferometry. Seismic interferometry, also called virtual source imaging (i.e., the real receivers can be treated as virtual sources), has drawn much attention in recent years \cite{nicolson_rayleigh_2014,galetti_transdimensional_2017} because it obviates the need for an active, controlled source by replacing it by a receiver at the desired location \cite{snieder_equivalence_2006}. We assume a homogeneous medium and disregard refraction of the waves, thus, the rays between pairs of source and receiver are straight. As shown in Fig.~\ref{fig1}, the slowness (i.e., reciprocal of velocity) map has been discretized as a $W_{1}\times W_{2}$ pixel image, and the receivers are randomly distributed on the 2-D slowness model. The slowness $\mathbf{s}$ can be written as the following linear model

\begin{equation}
\mathbf{s}=\mathbf{s}_{\mathrm{g}}+\mathbf{s}_{0} \in \mathbb{R}^{N},
\label{eq1}
\end{equation}

\noindent where $\mathbf{s}_{0}$ is reference slowness and $\mathbf{s}_{\mathrm{g}}$ is the perturbations from the reference, with $N=W_{1}\times W_{2}$. Furthermore, giving a tomography matrix $\mathbf{A}\in \mathbb{R}^{M \times N}$ with path lengths of $M$ rays, the formulation of observed traveltime $\mathbf{t}$ related to the tomography matrix $\mathbf{A}$ can be described as

\begin{equation}
    \mathbf{t}=\mathbf{A}\mathbf{s}=\mathbf{t}_{\mathrm{g}}+\mathbf{t}_{0} \in \mathbb{R}^{M},
    \label{eq2}
\end{equation}

\noindent where $\mathbf{t}_{\mathrm{g}}$ and $\mathbf{t}_{0}$ is the traveltime corresponding to the perturbations and the reference, respectively. Due to reference slowness $\mathbf{s}_{0}$ and tomography matrix $\mathbf{A}$ are given, the goal of traveltime tomography is to obtain perturbations $\mathrm{s}_{\mathrm{g}}$ from $\mathbf{t}_{\mathrm{g}}$ by inversion. The relationship between $\mathbf{t}_{\mathrm{g}}$ and $\mathrm{s}_{\mathrm{g}}$ can be expressed as

\begin{equation}
    \mathbf{t}_{\mathrm{g}}=\mathbf{A}\mathrm{s}_{\mathrm{g}} + \epsilon,
    \label{eq3}
\end{equation}

\noindent where $\epsilon \in \mathbb{R}^{M}$ is Gaussian noise $\mathcal{N}(\mathbf{0}, \sigma^{2}_{\epsilon}\mathbf{I})$, with mean $\mathbf{0}$ and covariance $\sigma^{2}_{\epsilon}\mathbf{I}$, and $\mathbf{I}$ is the identity matrix. According to Bayes's rule, we can obtain the posterior density of $\mathbf{s}_\mathrm{g}$ by 

\begin{equation}
    p\left(\mathbf{s}_{\mathrm{g}} \mid \mathbf{t}_\mathrm{g}\right) \propto p\left(\mathbf{t}_\mathrm{g} \mid \mathbf{s}_{\mathrm{g}}\right) p\left(\mathbf{t}_\mathrm{g}\right).
    \label{eq4}
\end{equation}

\noindent Here, we approximate $p\left(\mathbf{t}_\mathrm{g} \mid \mathbf{s}_{\mathrm{g}}\right)$ as Gaussian, thus it can be expressed as

\begin{equation}
    p\left(\mathbf{t}_\mathrm{g} \mid \mathbf{s}_{\mathrm{g}}\right) = \mathcal{N}\left(\mathbf{A} \mathbf{s}_\mathrm{g}, \boldsymbol{\Sigma}_{\epsilon}\right),
    \label{eq5}
\end{equation}

\noindent where $\boldsymbol{\Sigma}_{\epsilon} \in \mathbb{R}^{K \times K}$ is the covariance of the traveltime error. Taking the logarithm on both sides of Eq.~\ref{eq5} then we obtain

\begin{equation}
    \ln p\left(\mathbf{t}_\mathrm{g} \mid \mathbf{s}_{\mathrm{g}}\right) \propto-\frac{1}{2}\left(\mathbf{t}_\mathrm{g}-\mathbf{A} \mathbf{s}_{\mathrm{g}}\right)^{\mathrm{T}} \boldsymbol{\Sigma}_{\epsilon}^{-1}\left(\mathbf{t}_\mathrm{g}-\mathbf{A} \mathbf{s}_{\mathrm{g}}\right).
    \label{eq6}
\end{equation}

\noindent Taking the logarithm on both sides of Eq.~\ref{eq4} and substituting $\ln p\left(\mathbf{t}_\mathrm{g} \mid \mathbf{s}_{\mathrm{g}}\right)$ with Eq.~\ref{eq6}, we obtain

\begin{equation}
\begin{aligned}
    \ln p\left(\mathbf{s}_{\mathrm{g}} \mid \mathbf{t}_\mathrm{g}\right) &\propto \ln p\left(\mathbf{t}_\mathrm{g} \mid \mathbf{s}_{\mathrm{g}}\right) p\left(\mathbf{t}_\mathrm{g}\right)\\
    &\propto -\frac{1}{2}\left(\mathbf{t}_\mathrm{g}-\mathbf{A} \mathbf{s}_{\mathrm{g}}\right)^{\mathrm{T}} \boldsymbol{\Sigma}_{\epsilon}^{-1}\left(\mathbf{t}_\mathrm{g}-\mathbf{A} \mathbf{s}_{\mathrm{g}}\right) + \ln p(\mathbf{t}_\mathrm{g} ).
\end{aligned}
\label{eq7}
\end{equation}

\noindent Hence, we obtain the the Bayes maximum a posterior (MAP) objective, 

\begin{equation}
    \begin{aligned}
    \max \{ \ln p\left(\mathbf{s}_{\mathrm{g}} \mid \mathbf{t}_\mathrm{g}\right) \} &= \min \{-\ln p\left(\mathbf{s}_{\mathrm{g}} \mid \mathbf{t}_\mathrm{g}\right) \}\\
    &\propto \min \{\frac{1}{2}\left(\mathbf{t}_\mathrm{g}-\mathbf{A} \mathbf{s}_{\mathrm{g}}\right)^{\mathrm{T}} \boldsymbol{\Sigma}_{\epsilon}^{-1}\left(\mathbf{t}_\mathrm{g}-\mathbf{A} \mathbf{s}_{\mathrm{g}}\right) \}. 
    \end{aligned}
    \label{eq8}
\end{equation}

\noindent For simplicity, we assume the error is Gaussian independent and identically distributed (iid), i.e., $\Sigma_{\epsilon}=\sigma_{\epsilon}^{2}\mathbf{I}$. Furthermore, considering to constrain $\mathbf{s}_\mathrm{g}$ with regularization, Eq.~\ref{eq8} is thus

\begin{equation}
    \begin{aligned}
        \mathbf{s}^{\ast}_{\mathrm{g}} = \underset{\mathbf{s}_\mathrm{g}}{\arg \min}\left\{ \frac{1}{2\sigma^{2}_\epsilon}\left (\mathbf{t}_\mathrm{g}-\mathbf{A} \mathbf{s}_{\mathrm{g}}\right)^\mathrm{T} \left(\mathbf{t}_\mathrm{g}-\mathbf{A} \mathbf{s}_{\mathrm{g}}\right)\right \} \\ \text{subject to}\quad \eta\mathcal{R}\left(\mathbf{s}_\mathrm{g}\right),
    \end{aligned}
 \label{eq9}
\end{equation}

\noindent where $\eta$ denotes the weight, and $\mathcal{R}\left(\mathbf{s}_\mathrm{g}\right)$ denotes the regularization on $\mathbf{s}_\mathrm{g}$. To linearize this problem, we reformulate Eq.~\ref{eq9} as

\begin{equation}
    \mathbf{s}^{\ast}_{\mathrm{g}} = \underset{\mathbf{s}_\mathrm{g}}{\arg \min}\left\| \mathbf{t}_\mathrm{g}-\mathbf{A} \mathbf{s}_{\mathrm{g}} \right \|^{2}_{2} + \eta\mathcal{R}\left(\mathbf{s}_\mathrm{g}\right).
    \label{eq10}
\end{equation}

We adopt the LSQR \cite{rodgers_inverse_2008} to solve Eq.~\ref{eq10}, which regularizes the inversion with a global smoothing covariance. The estimated perturbations by LSQR can be written as

\begin{equation}
    \mathbf{s}^{\ast}_{\mathrm{g}} = \left( \mathbf{A}^{T}\mathbf{A} + \eta \boldsymbol{\Sigma}^{-1}_{\mathbf{L}} \right)^{-1} \mathbf{A}\left(\mathbf{t} - \mathbf{A}\mathbf{s}_{0} \right),
    \label{eq11}
\end{equation}

\noindent where $\boldsymbol{\Sigma}^{-1}_{\mathbf{L}}=\text{exp} (-D_{i,j}/L)$, with $D_{i,j}$ is the distance between cells $i$ and $j$ and $L$ is the smoothness length scale \cite{rodgers_inverse_2008, tarantola_inverse_2005}. 

Although we have obtained the estimated perturbations $\mathbf{s}^{\ast}_{\mathrm{g}}$, large difference probably still existed between $\mathbf{s}^{\ast}_{\mathrm{g}}$ and $\mathbf{s}_{\mathrm{g}}$ due to the linearization of LSQR. Therefore, we assume there is a mapping function $\mathcal{H}$ that can further minimize the gap between $\mathbf{s}^{\dagger}_{\mathrm{g}}=\mathcal{H}(\mathbf{s}^{\ast}_{\mathrm{g}}, \dots)$ and $\mathbf{s}_\mathrm{g}$. The true perturbations $\mathbf{s}_\mathrm{g}$ related to $\mathbf{s}^{\dagger}_{\mathrm{g}}$ can be expressed as

\begin{equation}
    \mathbf{s}_\mathrm{g} = \mathbf{s}^{\dagger}_{\mathrm{g}} + \tau,
    \label{eq12}
\end{equation}

\noindent where $\tau \in \mathbb{R}^{N}$ is Gaussian noise $\mathcal{N}(\mathbf{0}, \sigma^{2}_{\tau}\mathbf{I})$, with mean $\mathbf{0}$ and covariance $\sigma^{2}_{\tau}\mathbf{I}$. Similarly, we use Bayes's rule to derive the posterior density of $\mathbf{s}_\mathrm{g}$,

\begin{equation}
    p\left(\mathbf{s}_\mathrm{g} \mid \mathbf{s}^{\dagger}_{\mathrm{g}}\right) \propto 
    p\left(\mathbf{s}^{\dagger}_{\mathrm{g}} \mid \mathbf{s}_\mathrm{g}\right)p\left(\mathbf{s}^{\dagger}_{\mathrm{g}}\right).
    \label{eq13}
\end{equation}

\begin{figure*}[!tb]
\centering
\includegraphics[width=6.5in]{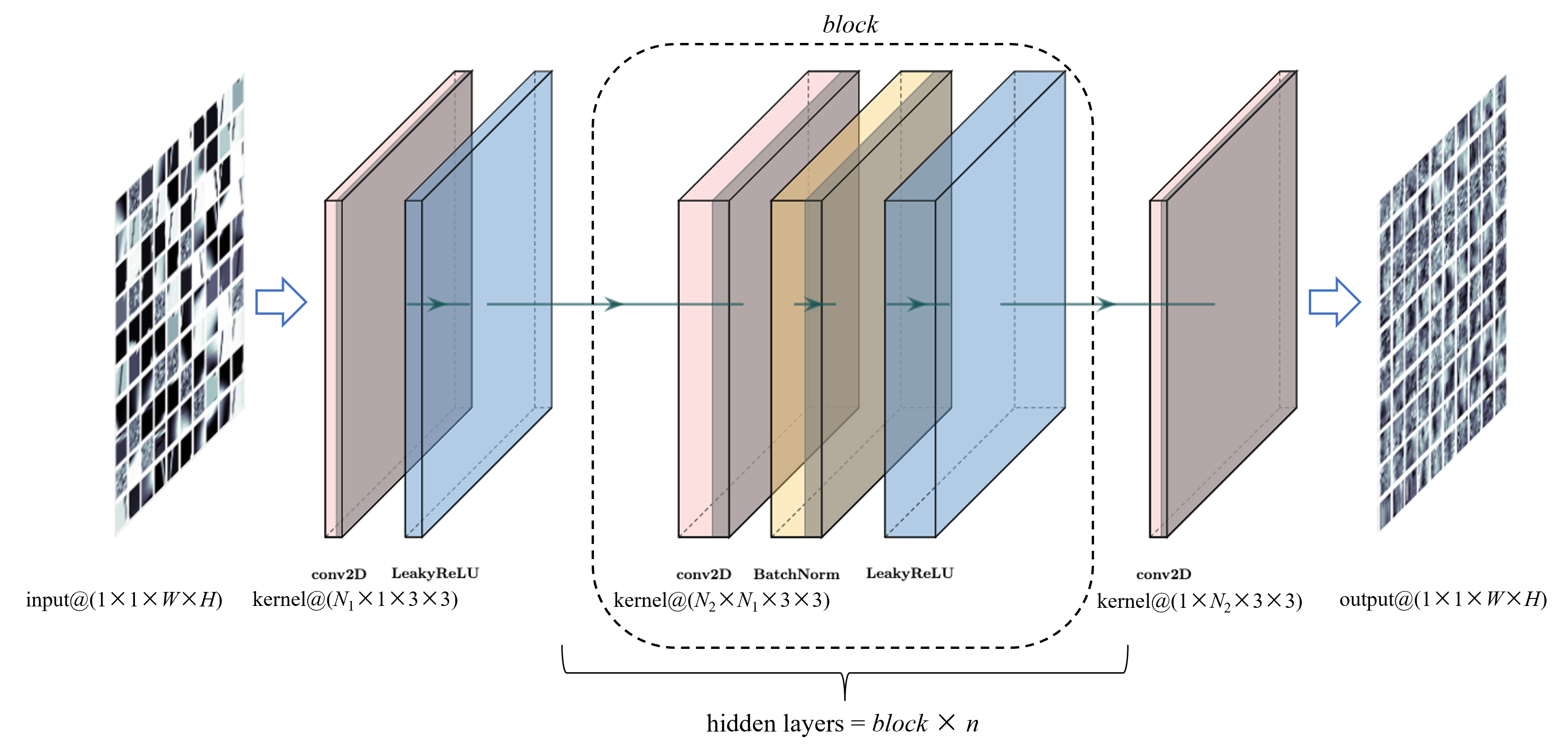}
\caption{ Illustration of our neural network. The input and output represent the initial and optimized dictionaries, respectively. Conv2D is a 2-D convolution layer, BatchNorm refers to batch normalization, and LeakyReLU is a nonlinear activation function. $n$ denotes the number of hidden layers.}
\label{nn}
\end{figure*}

\noindent Assuming the likelihood function $p\left(\mathbf{s}^{\dagger}_{\mathrm{g}} \mid \mathbf{s}_\mathrm{g}\right)$ to be a Gaussian distribution then it can be expressed by the following formula

\begin{equation}
    p\left(\mathbf{s}^{\dagger} \mid \mathbf{s}_\mathrm{g}\right) \propto \mathcal{N}\left(\mathbf{s}^{\dagger}_{\mathrm{g}}, \boldsymbol{\Sigma}_{\tau}\right),
    \label{eq14}
\end{equation}

\noindent where $\boldsymbol{\Sigma}_{\tau}$ represents the covariance of the perturbations error. Taking the logarithm on both sides of Eq.~\ref{eq14}, we can obtain 

\begin{equation}
    \begin{aligned}
        \ln p\left(\mathbf{s}_\mathrm{g} \mid \mathbf{s}^{\dagger}_{\mathrm{g}}\right) &\propto \ln p\left(\mathbf{s}^{\dagger}_{\mathrm{g}} \mid \mathbf{s}_\mathrm{g}\right)p\left(\mathbf{s}^{\dagger}_{\mathrm{g}}\right)\\ &\propto \frac{1}{2}\left(\mathbf{s}^{\dagger}_{\mathrm{g}}-\mathbf{s}_\mathrm{g}\right)^\mathrm{T}\boldsymbol{\Sigma}^{-1}_{\tau}\left(\mathbf{s}^{\dagger}_{\mathrm{g}}-\mathbf{s}_\mathrm{g}\right) + \ln p\left(\mathbf{s}^{\dagger}_{\mathrm{g}}\right).
    \end{aligned}
    \label{eq15}
\end{equation}

\noindent Consequently, we obtain the Bayes MAP objective with respect to $\mathbf{s}_\mathrm{g}$ and $\mathbf{s}^{\dagger}$ that is

\begin{equation}
    \begin{aligned}
        \max \left\{ \ln\left(\mathbf{s}_\mathrm{g} \mid \mathbf{s}^{\dagger}_{\mathrm{g}}\right) \right\} &= \min \left\{ -\ln\left(\mathbf{s}_\mathrm{g} \mid \mathbf{s}^{\dagger}_{\mathrm{g}}\right) \right\}\\
        &\propto \min \left\{ \frac{1}{2}\left(\mathbf{s}^{\dagger}_{\mathrm{g}}-\mathbf{s}_\mathrm{g}\right)^\mathrm{T}\boldsymbol{\Sigma}^{-1}_{\tau}\left(\mathbf{s}^{\dagger}_{\mathrm{g}}-\mathbf{s}_\mathrm{g}\right)\right\}.
    \end{aligned}
    \label{eq16}
\end{equation}

\noindent For simplicity, we often assume $\boldsymbol{\Sigma}_{\tau}$ are Gaussian iid, Eq.~\ref{eq16} thus becomes

\begin{figure*}[!tb]
\centering
\includegraphics[width=4in]{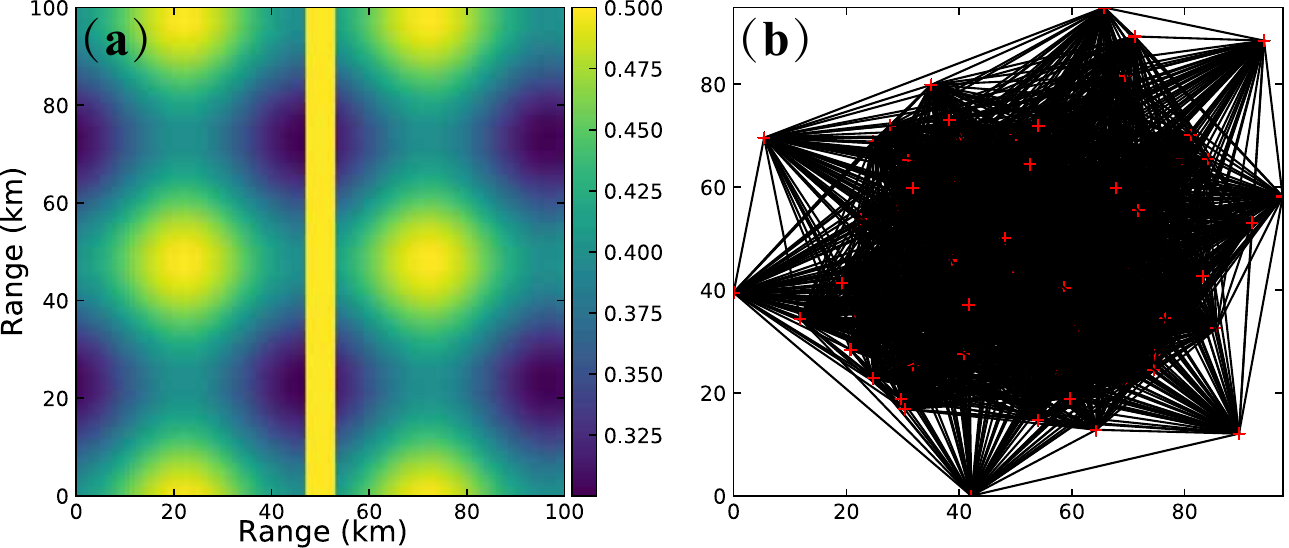}
\caption{(a) Synthetic slowness map with dimensions of $W_1=W_2=100$ pixels (1 km/pixel). (b) Ray sampling with 64 receivers (red crosses).}
\label{model}
\end{figure*}

\begin{equation}
    \mathbf{s}^{\ddagger}_{\mathrm{g}} = \underset{\mathbf{s}^{\dagger}_{\mathrm{g}}}{\arg \min }\left\{ \frac{1}{{2\sigma}^{2}_{\tau}}\left\|\mathbf{s}^{\dagger}_{\mathrm{g}}-\mathbf{s}_{\mathrm{g}}\right\|_{2}^{2}\right\}.
    \label{eq17}
\end{equation}

\subsection{Solving the MAP}
To solve Eq.~\ref{eq17}, we propose a two-step strategy: (1) \emph{warming up}. The \emph{warming up} phrase can provide the initial dictionary by the dictionary learning method (i.e., iterative thresholding and signed K-means (ITKM) \cite{schnass_local_2015}) and the initial code through the sparse coding algorithm (orthogonal matching pursuit (OMP) \cite{pati_orthogonal_1993}); (2) \emph{dictionary optimization}. The \emph{dictionary optimization} phrase is used to optimize the initial dictionary through the shallow and simple NN, and then we reconstruct the high-resolution VMs through the optimized dictionary and the updated code. The illustration of our method is shown in Fig.~\ref{method}.

\subsubsection{Warming up}
Dictionary learning is a kind of data-driven approach, which can reconstruct signal using a small number of vectors, called \emph{atoms}, from an overcomplete matrix. Inspired by the performance of dictionary learning in traveltime tomography \cite{bianco_travel_2018, zhou_high-resolution_2022}, we adopt dictionary learning and sparse coding to complete the task in this section. We train a dictionary from the patches of the estimation by LSQR as the initial dictionary and compute the initial code corresponding to the initial dictionary through sparse coding. The relationship of dictionary $\mathbf{D}$, code $\mathbf{X}$ and these patches can be expressed as 

\begin{equation}
    \mathbf{Y}_{i}=\mathbf{R}_{i}\mathbf{s}^{\ast}_{\mathrm{g}} \approx
    \mathbf{D}\mathbf{X}_{i},  
    \label{eq18}
\end{equation}

\noindent where $\mathbf{Y}_{i}\in \mathbb{R}^{\sqrt{n} \times \sqrt{n}}$ represents the \emph{i-th} patch sampled from $\mathbf{s}^{\ast}_{\mathrm{g}}$ by the binary matrix $\mathbf{R} \in \{0, 1\}$ (Fig.~\ref{fig1}). Using ITKM algorithm to learn the initial dictionary $\mathbf{D}$,

\begin{equation}
    \mathbf{D}=\mathbf{ITKM}(\mathbf{Y}, n_a, T),
    \label{eq19}
\end{equation}

\noindent where $\mathbf{D} \in \mathbb{R}^{j \times n_a}$, $j=\sqrt{n} \times \sqrt{n}$ is the length of atom, $n_a$ is the number of atoms, and $T$ is the sparse level which is used to remain $T$ largest values of $\mathbf{D}_{i}$. The smaller the $T$, the higher the sparse level. We then compute the initial code $\mathbf{X} \in \mathbb{R}^{n_a \times \sqrt{n}}$ by OMP method,

\begin{equation}
    \mathbf{X} = \mathbf{OMP}(\mathbf{D}, \mathbf{Y}, H_0),
    \label{eq20}
\end{equation}

\noindent where $H_0$ represents the sparsity level for $\mathbf{X}_{i}$.

\subsubsection{Dictionary optimization}
After achieving the initial dictionary and the initial code, we take the NN as the mapping function $\mathcal{H}$, i.e., $\mathcal{H}(\cdot) := \mathcal{NN}\left(\cdot ;\boldsymbol{\theta} \right)$, where $\boldsymbol{\theta}$ are the parameters of NN. Therefore, we can obtain the optimized dictionary $\mathbf{D}^{\dagger}$ by

\begin{equation}
    \mathbf{D}^{\dagger}=\mathcal{NN}\left(\mathbf{D} ;\boldsymbol{\theta} \right),
    \label{eq21}
\end{equation}

\noindent and the updated code $\mathbf{X}^{\dagger}$ is related to $\mathbf{D}^{\dagger}$ by

\begin{equation}
    \mathbf{X}^{\dagger}=\mathbf{OMP}(\mathbf{D}^{\dagger}, \mathbf{Y}, H_1),
    \label{eq22}
\end{equation}

\noindent where $H_1$ is the sparsity level for $\mathbf{X}^{\dagger}_{i}$. Using $\mathbf{D}^{\dagger}$ and $\mathbf{X}^{\dagger}$ to reconstruct $\mathbf{s}^{\ast}_{\mathrm{g}}$, and substituting them into Eq.~\ref{eq17}, we obtain 

\begin{equation}
    \mathbf{s}^{\ddagger}_{\mathrm{g}} = \underset{\mathbf{D}^{\dagger}\mathbf{X}^{\dagger}}{\arg \min }\left\{ \sum_{i}\left\|\mathbf{D}^{\dagger}\mathbf{X}^{\dagger}_{i}-\mathbf{R}_{i}{\mathbf{s}_\mathrm{g}}\right\|_{2}^{2}\right\}.
    \label{eq23}
\end{equation}

\noindent Differentiating Eq.~\ref{eq23}, we can obtain

\begin{equation}
    \begin{aligned}
        \frac{d}{d\mathbf{s}_{\mathrm{g}}}&\left\{ \sum_{i}\left\|\mathbf{D}^{\dagger}\mathbf{X}^{\dagger}_{i}-\mathbf{R}_{i}{\mathbf{s}_\mathrm{g}}\right\|_{2}^{2}\right\} \\
        &=2j\mathbf{I}\mathbf{s}_\mathrm{g}-2\sum_{i}\mathbf{R}_{i}^{T}\mathbf{D}^{\dagger}\mathbf{X}^{\dagger},
        \label{eq24}
    \end{aligned}
\end{equation}

\noindent where $j\mathbf{I}=\sum_{i}\mathbf{R}_{i}^{T}\mathbf{R}_{i}$. What is more, due to the patches are centered \cite{mairal_sparse_2014}, i.e., the mean of patch \emph{i} is subtracted, we add the mean of each patch back into reconstructed patch before computing $\mathbf{s}^{\ddagger}_{\mathrm{g}}$ by

\begin{equation}
    \mathbf{D}^{\dagger}\mathbf{X}^{\dagger} \gets \mathbf{D}^{\dagger}\mathbf{X}^{\dagger}+\overline{\mathbf{Y}}
\end{equation}

\noindent where $\overline{\mathbf{Y}}$ is the mean of the patches. More details can be found in \cite{bianco_travel_2018}. Consequently, we can derive

\begin{equation}
    \mathbf{s}^{\ddagger}_{\mathrm{g}}=\frac{1}{j}\sum_{i}\mathbf{R}_{i}^{T}\mathbf{D}^{\dagger}\mathbf{X}^{\dagger}.
    \label{eq25}
\end{equation}

\noindent In addition to the reference $\mathbf{s}_{0}$ and the perturbations $\mathbf{s}^{\ddagger}_{\mathrm{g}}$, we also add the perturbations $\mathbf{s}^{\ast}_{\mathrm{g}}$ estimated by LSQR into the slowness map used for the final interpretation by

\begin{equation}
    \mathbf{s} = \alpha\mathbf{s}_{0} + \beta \mathbf{s}^{\ast}_{\mathrm{g}} + \gamma\mathbf{s}^{\ddagger}_{\mathrm{g}},
    \label{26}
\end{equation}

\noindent where the $\alpha, \beta, \gamma \in [0, 1]$ denote weights.

\subsubsection{NN designing and training}
In this section, we design a \emph{shallow} and \emph{simple} NN instead of using the deep and complex NN in many deep-learning-based tomographic methods. As shown in Fig.\ref{nn}, plot using the PlotNeuralNet package (\url{https://github.com/HarisIqbal88/PlotNeuralNet}), our NN consists of 2-D convolution, batch normalization and LeakyReLU layer. The first convolution layer is followed by a LeakyReLU layer and the block that is composed of convolution, batch normalization and LeakyReLu. The size of filter kernels of the first convolution layer is $64\times1\times3\times3$, with the format of $\text{number of filters} \times \text{number of channels} \times \text{width} \times \text{height}$. From the second to the penultimate convolution layer, we set the size of all filter kernels to $64\times 64\times 3 \times 3$. For the last convolution layer, the size of filter kernels is set to $64\times 1\times 3 \times 3$. The LeakyReLU is a popularly used non-linear activation function to introduce non-linearity to NNs, and it is defined by 

\begin{figure*}[!tb]
\centering
\includegraphics[width=7in]{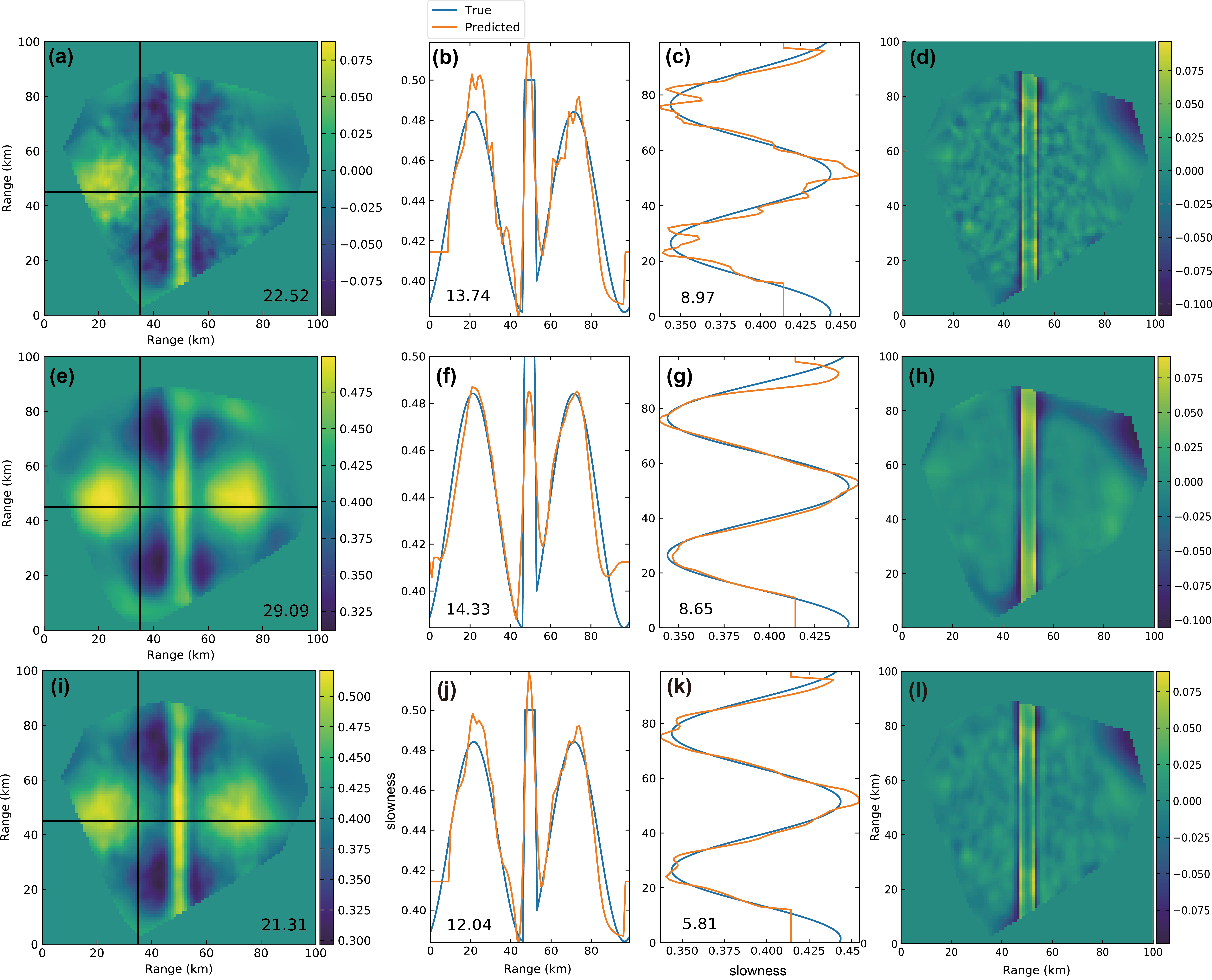}
\caption{Comparison of 2-D slowness, 1-D slowness (from the black lines in 2-D slowness) against true slowness (Fig.~\ref{model}(a)), and slowness errors by LSQR, dictionary learning, and the proposed method ($\sigma=0.02$). (a)-(d) Results by LSQR. (e)-(h) Results by dictionary learning with 150 atoms. (i)-(l) Results by our method with 150 atoms. RMSE values are printed on these slowness maps.}
\label{reslut1}
\end{figure*}

\begin{equation}
    \text{LeakyReLU}(x)=\max(0,x)+\phi \ast \min(0,x).
    \label{eq27}
\end{equation}

\noindent We set $\phi$ to 0.01 in the next section of numerical tests.

To train this NN in a label-free manner, we use the initial dictionary obtained in the \emph{warming up} step and the observed traveltime $\mathbf{t}$ as the training data. Hence, the input for the NN training is unique, and the NNs' prediction is the optimized dictionary that will be used to reconstruct the VMs with high-resolution. We iteratively perform the training to minimize the traveltime misfit that can be defined as

\begin{equation}
    \mathcal{L}(\mathbf{s}^{\ddagger}, \mathbf{s}_{0},\mathbf{A}, \mathbf{V}, \mathbf{t}) = \frac{1}{M}\left \| \mathbf{A} \left(\mathbf{s}^{\ddagger} \odot \mathbf{V} + \mathbf{s}_{0}\right)-\mathbf{t}\right\|_{2}^{2},
    \label{eq28}
\end{equation}

\noindent where $\mathbf{V}$ is the binary mask for the region covered by rays, $\odot$ denotes Hadamard product, and $M$ represents the number of elements of $\mathbf{t}$. This loss function measures the mean square error (MSE) between the traveltime of the inverted slowness and the observed traveltime for the area covered by rays. Once the training is completed, the output of the last training epoch will be directly used to compute the code $ \mathbf{X}^{\dagger}$ by Eq.~\ref{eq22}, and then we obtain perturbations $\mathbf{s}^{\ddagger}$ through Eq.~\ref{eq25}. The implementation detail of NN training is summarized in Algorithm \ref{alg}. 

\begin{algorithm}
\caption{NN training strategy}
\label{alg}
\begin{algorithmic}
%   \Require $\mathbf{D}_{0}$: initial dictionary; $\mathbf{X}$: initial code; $\mathcal{NN}$: pre-designed neural NN; $\mathbf{R}$: binary mask for patch sampling; $epoch$: the number of iteration for NN training; $\mathbf{V}$: binary mask for the region covered by rays; $\mathbf{t}$: traveltime; $\mathbf{s}_{0}$: reference slowness 
   \Ensure optimal $\mathbf{D}^{\dagger}$
    \State initial $n=1$ and initialize the weights $\boldsymbol{\theta}$ of $\mathcal{NN}$ with uniform distribution
    \While{$n <= epoch$}
          \State compute prediction $\mathbf{D}^{\dagger}=\mathcal{NN}(\mathbf{D}_{0}; \boldsymbol{\theta})$
          \State compute perturbations $    \mathbf{s}^{\ddagger}=\frac{1}{j}\sum_{i}\mathbf{R}_{i}^{T}\left(\mathbf{D}^{\dagger}\mathbf{X}+\overline{\mathbf{Y}}\right)$
          \State compute loss using Eq.~\ref{eq28} 
          \State update $\boldsymbol{\theta}$
          \State $n \gets n+1$
    \EndWhile
\end{algorithmic}
\end{algorithm}

\begin{figure}[!tb]
\centering
\includegraphics[width=3.5in]{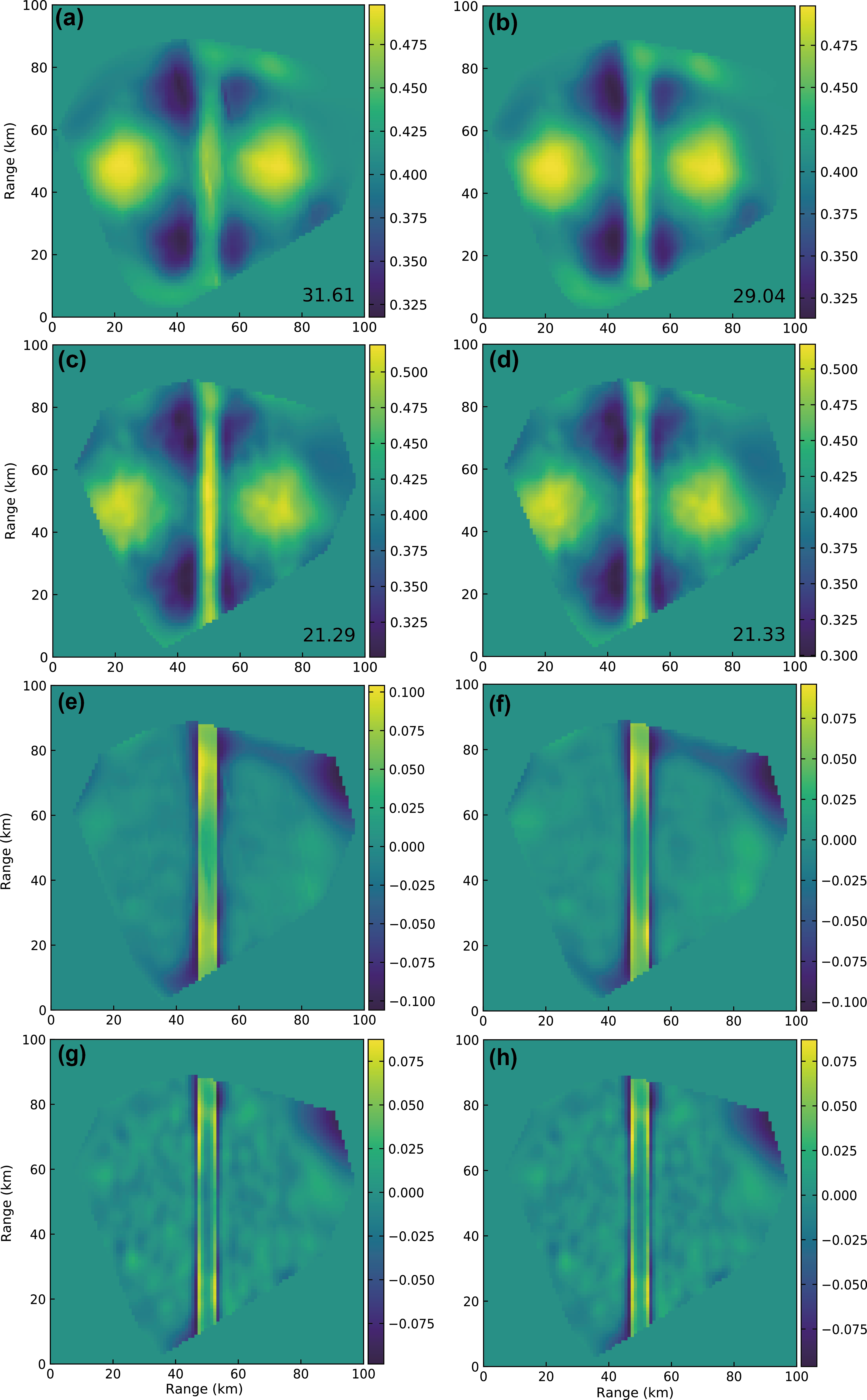}
\caption{Slowness and slowness errors by dictionary learning and the proposed method ($\sigma=0.02$). (a)-(b) Results by dictionary learning with 50 and 100 atoms. (c)-(d) Results byv our method results with 50 and 100 atoms. (e)-(h) Slowness errors corresponding to (a)-(d). RMSE values are printed on these slowness maps.}
\label{reslut2}
\end{figure}

\section{Numerical tests}
In this section, we test the effectiveness of our methods using two velocity models: the smooth-discontinuous model and the Marmousi model. It should to note that this paper focuses on enhancing the resolution of VMs by traditional tomography. As such, we do not demonstrate inversion results for $\sigma=0$, since many traditional methods can already achieve satisfactory results in noise-free cases.

\subsection{Smooth-discontinuous model}
The smooth-discontinuous model is a 2-D synthetic slowness map (Fig.~\ref{model}a) which has 100 pixels (km) in both vertical and horizontal direction. On this map, 64 receivers are randomly distributed on the smooth-discontinuous map, and 2016 straight rays go through it (Fig.~\ref{model}b). We test their performance of the model on the traveltime with Gaussian noise of standard deviation $\sigma=0.02$ and $\sigma=0.05$, respectively. 

Here, we only compare the proposed method with LSQR and dictionary learning for the following two reasons: 1) Our approach aims to improve the low-resolution VMs using traditional tomography methods such as LSQR; and 2) the proposed method combines NN and dictionary learning, meaning that LSQR and/or dictionary learning can be substituted with other algorithms such as total variation. The implementation details of LSQR \cite{rodgers_inverse_2008} are described in Eq.~\ref{eq11}, while the dictionary-learning tomography is implemented by performing LSQR, dictionary learning, and sparse coding iteratively to solve Eq.~\ref{eq10} (see Algorithm \ref{alg2}).

\begin{algorithm}
\caption{Dictionary-learning tomography}
\label{alg2}
\begin{algorithmic}
%   \Require $\mathbf{R}$: binary mask for patch sampling; $k$: the number of iteration; $\mathbf{t}$: traveltime; $\mathbf{s}_{0}$: reference slowness
   \Ensure optimal $\mathbf{D}$ and $\mathbf{X}$
    \State initial $n=1$ and $\mathbf{s}_\mathrm{g}=0$
    \While{$n <= k$}
          \State $\text{dt}=\mathbf{t}-\mathbf{A}(\mathbf{s}_\mathrm{g} + \mathbf{s}_{0})$
          \State $\text{ds}=\textbf{LSQR}(\mathbf{A},\text{dt}, \text{damp}, \text{iter})$
          \State $\mathbf{s} = \text{ds} + \mathbf{s}_{0}$
          \State $\mathbf{Y}=\mathbf{R}\mathbf{s}$
          \State  $\mathbf{D}=\mathbf{ITKM}(\mathbf{Y}, T_d)$ 
          \State $\mathbf{X} = \mathbf{OMP}(\mathbf{D}, \mathbf{Y}, H_d)$ 
          \State $\mathbf{s}_{\mathrm{g}}=\frac{1}{j}\sum_{i}\mathbf{R}_{i}^{T}\left(\mathbf{D}\mathbf{X} + \overline{\mathbf{Y}}\right)$
          \State $n \gets n+1$
    \EndWhile
\end{algorithmic}
\end{algorithm}

We keep some hyper-parameters the same for both $\sigma=0.02$ and $\sigma=0.05$. In LSQR, we set the $\eta$ and $L$ (Eq.~\ref{eq11}) to 10 $\text{km}^2$ and 20 $\text{km}$ respectively, and set the initial velocity to a constant. In dictionary learning, we assign a damping coefficient $=10$, a patch size of $10\times10$, and 1000 iterations for LSQR. The iteration $k$ is set to 50. Spares level $T_d$ and $H_d$ are both set to 2 in Algorithm~\ref{alg2}. Patches with more than 10\% of pixels not sampled by rays are excluded from dictionary training \cite{bianco_travel_2018}. In the proposed method, the NN contains five convolution layers, three batch normalization layers, and four LeakyReLU layers (i.e., $n=3$ in Fig.~\ref{nn}). The filter kernel settings are described in the \emph{NN designing and training} section. The NN training epoch is set to 50 on the PyTorch platform using the AdamW algorithm with a learning rate of 0.001. We use sparse levels $T=1$ (Eq.~\ref{eq19}) and $H_0=1$ (Eq.~\ref{eq20}) in the \emph{warming up}. The patch size in the proposed method is set to $20\times20$ pixels.

We adopt root mean squared error (RMSE) to quantify the quality of inversion results, the RMSE (ms/km) is expressed as

\begin{equation}
    \text{RMSE}=\sqrt{\frac{1}{NP} \sum_{\mathrm{n}}^{N} \sum_{p}^{P}\left(s_{n, p}\mathbf{V}-s_{n,p}^{\prime}\mathbf{V}\right)^{2}} \times 1000,
    \label{eq29}
\end{equation}

\noindent where $s$ and $s^{\prime}$ denote the true and estimated slowness, respectively. 

For $\sigma=0.02$, we set $H_1$ (Eq.~\ref{eq22}) to 25 to reconstruct the perturbations. From the inverse results (Fig.~\ref{reslut1}), it can be seen that the slowness map produced by LSQR still contains a lot of noise. Although dictionary learning produces a smooth result, it is unable to effectively invert the slowness of the discontinuous region between approximately 40 to 60 km. In comparison with LSQR, the slowness map produced by the proposed method has less noise and higher fidelity, demonstrating its effectiveness in improving the resolution of VMs by LSQR. The 1-D slowness profiles show that the proposed method can smooth the signal and minimize the gap between improved and true slowness. Dictionary learning produces smoother slowness profiles and also smooths the anomaly boundaries. Slowness errors further illustrate that the VMs produced by the proposed method have less noise compared to the errors produced by LSQR, while dictionary learning produces more significant errors at discontinuous regions and results higher RMSE.  

\begin{figure*}[!tb]
\centering
\includegraphics[width=7in]{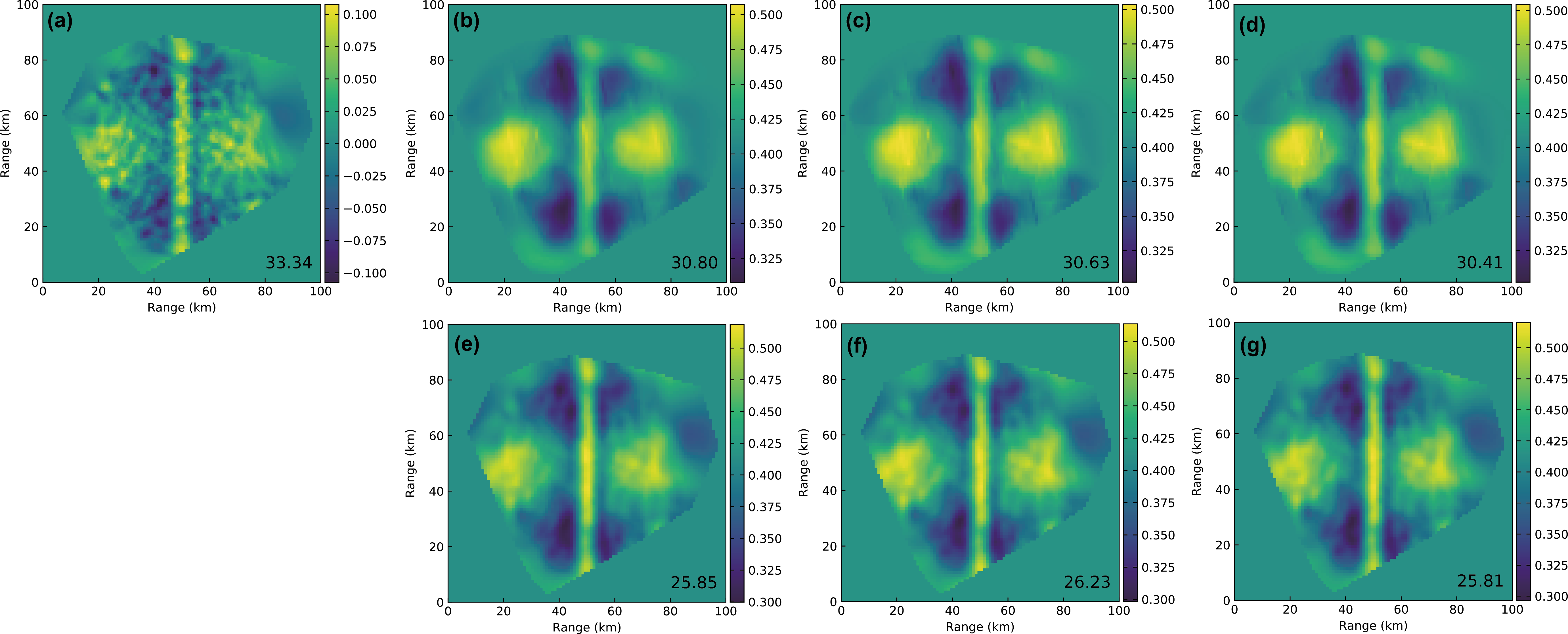}
\caption{Slowness maps by LSQR, dictionary learning, and the proposed method ($\sigma=0.05$). (a) Results by LSQR. (b)-(d) Results by dictionary learning with 50, 100, and 150 atoms. (e)-(g) Results by our method with 50, 100, and 150 atoms. RMSE values are printed on these slowness maps.}
\label{reslut3}
\end{figure*}

\begin{figure*}[!tb]
\centering
\includegraphics[width=7in]{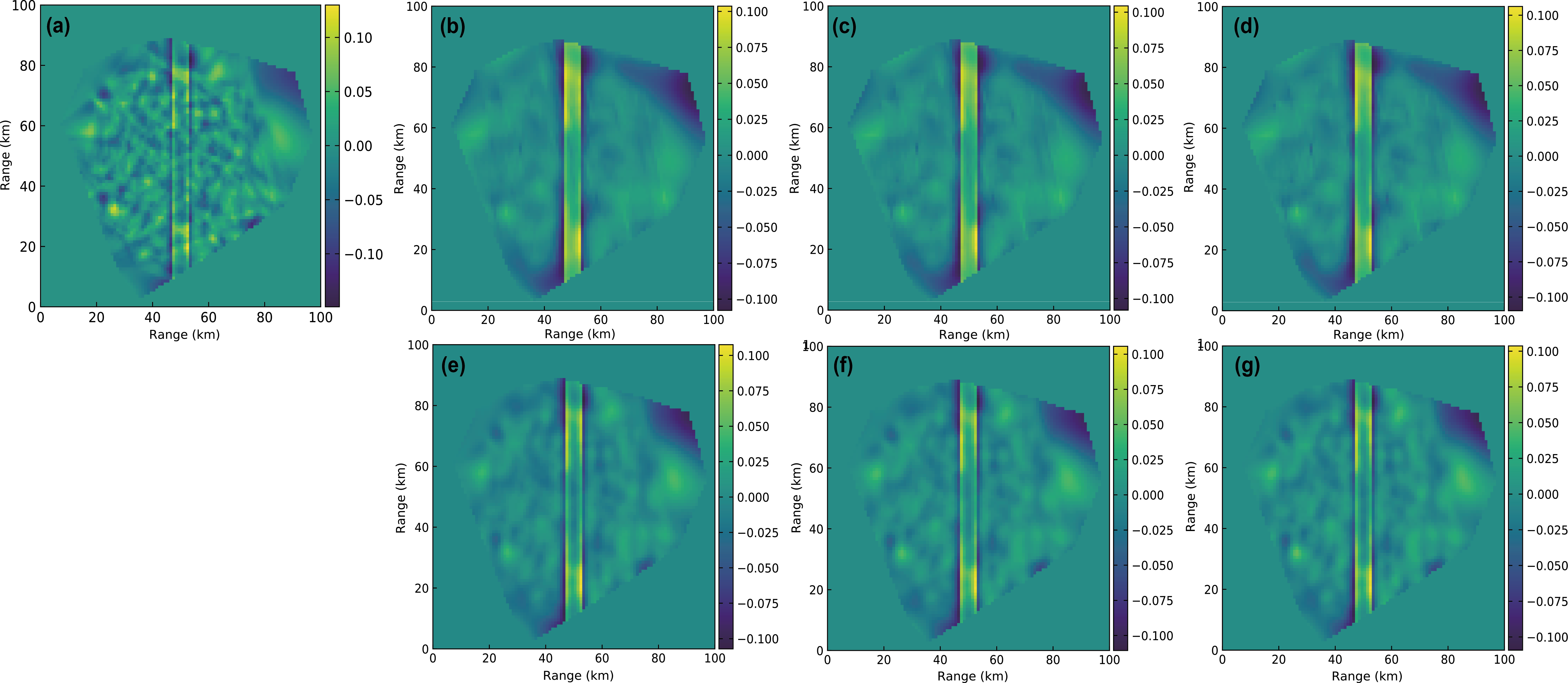}
\caption{Slowness errors by LSQR, dictionary learning, and the proposed method ($\sigma=0.05$). (a) Results by LSQR. (b)-(d) Results by dictionary learning results with 50, 100, and 150 atoms. (e)-(g) Results by our method with 50, 100, and 150 atoms.}
\label{reslut4}
\end{figure*}

The performance of dictionary learning heavily depends on the number of atoms. This is the reason why dictionary learning usually requires an over-complete dictionary. However, more atoms mean higher computational cost. To compare the influence of the number of atoms on dictionary learning tomography and the proposed method, we reduced the number of atoms to 50 and 100. As the number of atoms decreases, the slowness maps obtained by dictionary learning become smoother and their resolution is significantly reduced (Fig.~\ref{reslut2}(a) and (b)). In contrast to dictionary learning, the resolution of slowness by the proposed method remains almost unchanged (Fig.~\ref{reslut2}(c) and (d)) compared with Fig.~\ref{reslut1}(i). The slowness errors (Fig.~\ref{reslut2}(e)-(h)) demonstrate that there are obvious errors in the discontinuous area of results obtained by dictionary learning, indicating that decreasing the number of atoms reduces its effectiveness.

\begin{figure*}[]
\centering
\includegraphics[width=6in]{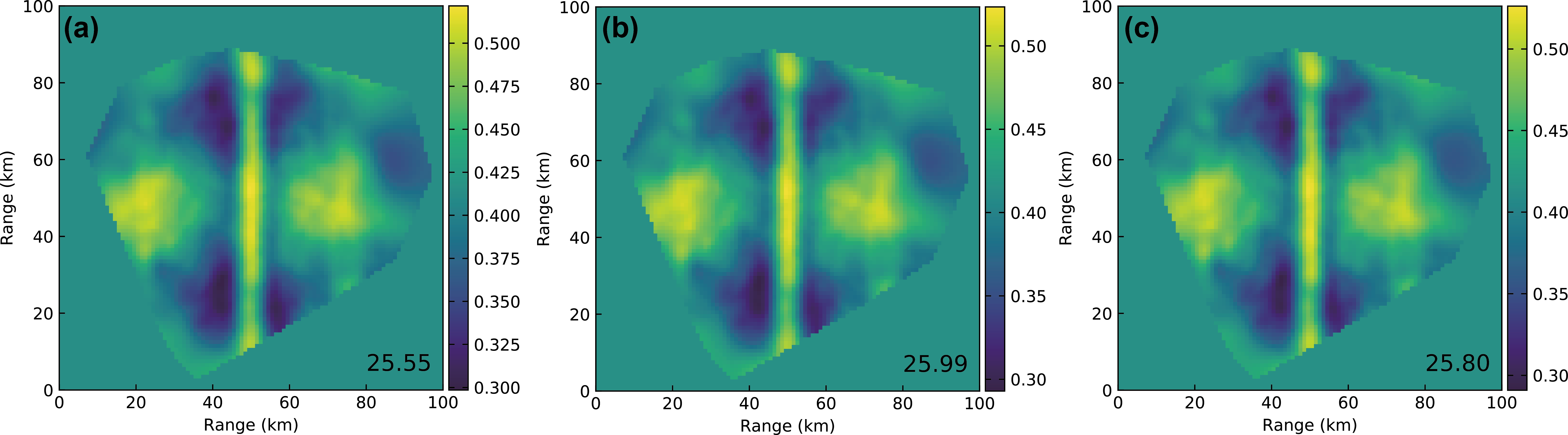}
\caption{Slowness results obtained by the trained neural network with $\sigma=0.02$ and 50 atoms for $\sigma=0.05$. (a)-(c) Results with 50, 100, and 150 atoms. RMSE values are printed on these slowness maps.}
\label{gener}
\end{figure*}

\begin{figure*}[]
\centering
\includegraphics[width=6in]{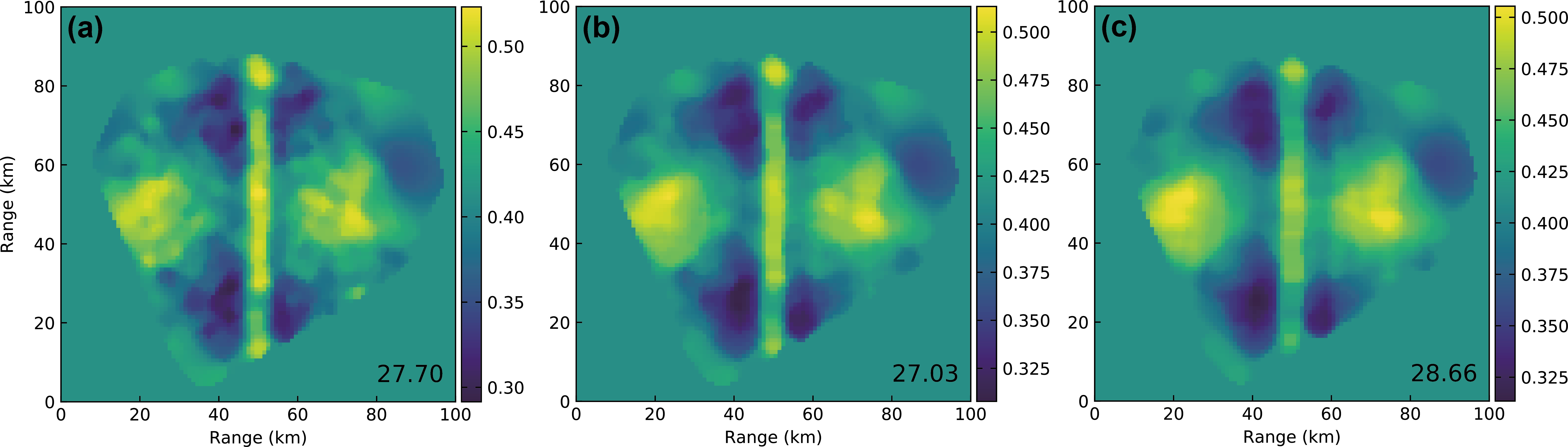}
\caption{Slowness results obtained by the median filter method ($\sigma=0.05$). (a) Result by filter size of $5\times5$. (b) Result by filter size of $7\times7$. (c) Result by filter size of $9\times9$. RMSE values are printed on these slowness maps.}
\label{media}
\end{figure*}

\begin{table*}[]
\renewcommand{\arraystretch}{1.5}
\centering
\caption{Comparsion of LSQR, dictionary learning and the proposed method tomography RMSE (ms/km).}
\label{tab}
\begin{tabular}{|l|cccc|cccc|}
\hline
noise level  & \multicolumn{4}{c|}{$\sigma=0.02$}                                                                                               & \multicolumn{4}{c|}{$\sigma=0.05$}                                                                                               \\ \hline
number of atoms  & \multicolumn{1}{c|}{-}     & \multicolumn{1}{c|}{50}             & \multicolumn{1}{c|}{100}            & 150            & \multicolumn{1}{c|}{-}     & \multicolumn{1}{c|}{50}             & \multicolumn{1}{c|}{100}            & 150            \\ \hline
LSQR                 & \multicolumn{1}{c|}{22.52} & \multicolumn{1}{c|}{-}              & \multicolumn{1}{c|}{-}              & -              & \multicolumn{1}{c|}{33.34} & \multicolumn{1}{c|}{-}              & \multicolumn{1}{c|}{-}              & -              \\ \hline
dictionary Learning          & \multicolumn{1}{c|}{-}     & \multicolumn{1}{c|}{31.61}          & \multicolumn{1}{c|}{29.04}          & 29.09          & \multicolumn{1}{c|}{-}     & \multicolumn{1}{c|}{30.80}          & \multicolumn{1}{c|}{30.63}          & 30.41          \\ \hline
the proposed method                        & \multicolumn{1}{c|}{-}     & \multicolumn{1}{c|}{\textbf{21.29}} & \multicolumn{1}{c|}{\textbf{21.33}} & \textbf{21.31} & \multicolumn{1}{c|}{-}     & \multicolumn{1}{c|}{\textbf{25.85}} & \multicolumn{1}{c|}{\textbf{26.23}} & \textbf{25.81} \\ \hline
\end{tabular}
\end{table*}

\begin{figure}[]
\centering
\includegraphics[width=3in]{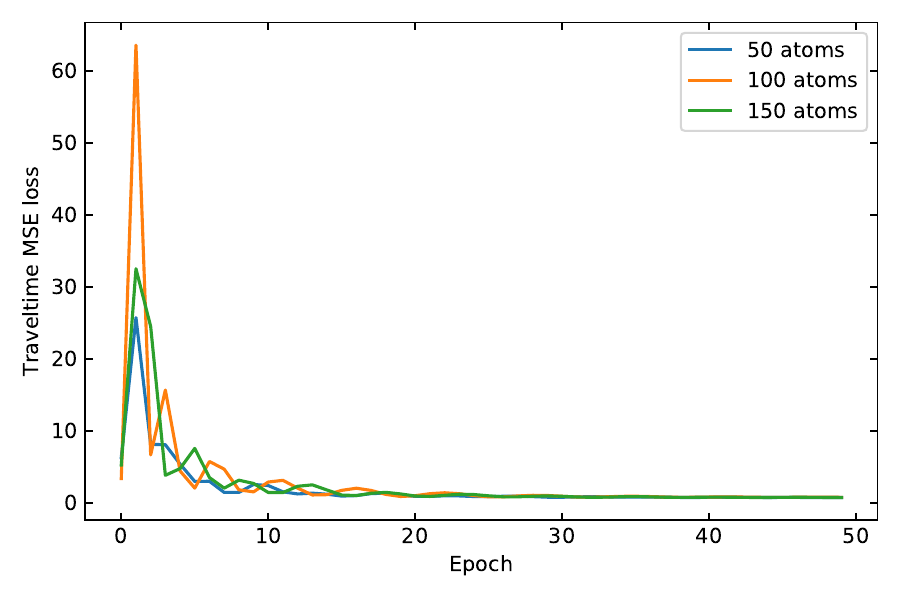}
\caption{Traveltime MSE (s/km) loss v.s. epoch of NN training with different numbers of atoms ($\sigma=0.05$).}
\label{train_loss}
\end{figure}

To further evaluate the effectiveness of the proposed method in the presence of stronger noise, we test its performance on the traveltime with with a noise level of $\sigma=0.05$. In this test, only $H_1$ (Eq.~\ref{eq22}) is set to 5 in the dictionary optimizing step to account for the stronger impact of noise on inversion results. All other hyper-parameters used in competing methods and the proposed method remained the same as in the previous test where $\sigma=0.02$. As shown in Fig.~\ref{reslut3}(a), the resolution of LSQR inversion decreased dramatically and anomaly shapes became chaotic. Dictionary learning produces very smooth results with low resolution, especially at the anomaly boundaries (Fig.~\ref{reslut3}(b)-(d)). Fig.~\ref{reslut3}(e)-(g) demonstrate that the proposed method can still successfully improve resolution even when the resolution of LSQR's estimation very low. Slowness errors (Fig.~\ref{reslut4}) further reveal that LSQR is sensitive to noise while dictionary learning suppresses much noise but sacrifices detail at the anomaly boundaries and discontinuous regions. The slowness errors of the proposed method have less noise compared to LSQR and fewer errors in the discontinuous regions compared to dictionary learning, suggesting a trade-off between smoothness and resolution of slowness maps. Traveltime RMSEs for all three approaches are listed in Table \ref{tab}, and we can clearly observe that the proposed method achieves achieves the lowest RMSE in each test, proving its robustness against different noise levels and atom numbers. As shown in Fig.~\ref{train_loss}, the training loss curves for traveltime MSE decreased rapidly and converged after approximately 10 epochs.

In addition, to test the generalization of the proposed method, we apply the well-trained neural network to the traveltime with higher noise levels and varying numbers of atoms. We apply the neural network used for the estimation with $\sigma=0.02$ and 50 atoms to improve the estimation with $\sigma=0.05$ (Fig.~\ref{reslut3}a). In this test, all hyperparameters were identical to those in the previous experiment with $\sigma=0.05$ except the NN model. The results show that slowness (Fig.~\ref{gener}) was effectively improved with different numbers of atoms and that the resolution and RMSEs of these velocity models were very similar to those in the experiment with $\sigma=0.05$ (Fig.~\ref{reslut3}e-g). This indicates that the proposed method has good generalization capabilities and that we can use trained neural networks to further improve computational efficiency for the same inversion tasks.

To demonstrate the difference between our method and filter-based methods, we further investigate the performance of filter-based methods in improving the resolution of VMs using LSQR. We adopt a commonly used filter-based method-the median filter with different filter sizes to improve LSQR’s estimation. As shown in Fig.~\ref{media}, the slowness maps obtained using the median filter resemble abstract paintings, and the larger the filter size, the fewer high frequencies are present.

\subsection{Marmousi model}
\begin{figure}[]
\centering
\includegraphics[width=3.3in]{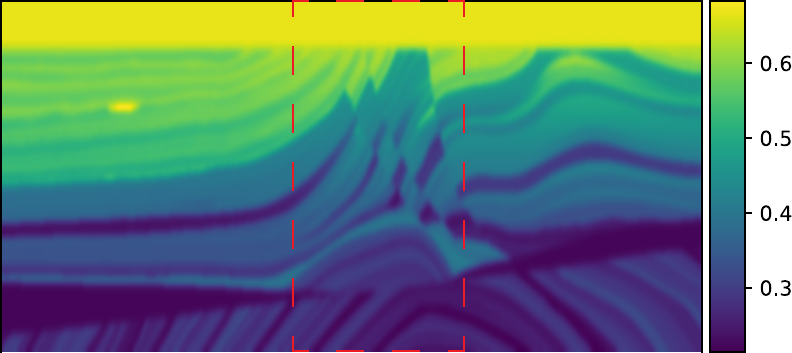}
\caption{Marmousi model (s/km).}
\label{marmousi}
\end{figure}

\begin{figure*}[]
\centering
\includegraphics[width=6in]{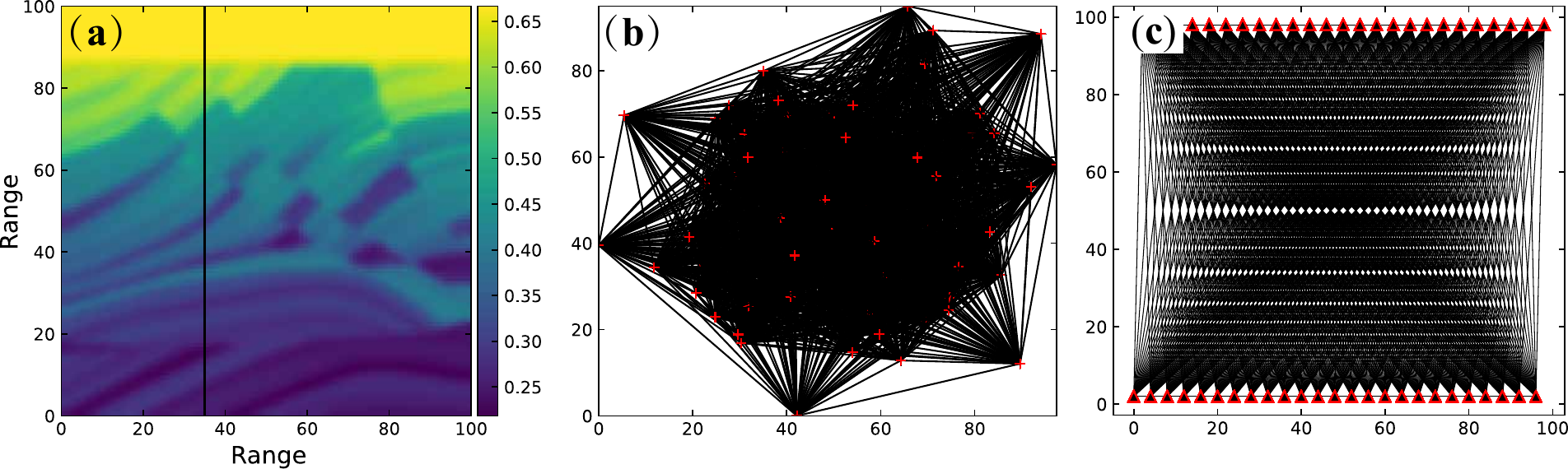}
\caption{(a) Central part of Marmousi model. (b) Ray sampling with 64 random regularly distributed (red crosses). (c) Ray sampling with 100 regularly distributed receivers (red triangles).}
\label{marmousi_local}
\end{figure*}

We further test the performance of our method using the most heterogeneous part (Fig.~\ref{marmousi_local}a) of the smoothed Marmousi model (Fig.~\ref{marmousi}) which is smoothed using a Gaussian filter. We obtain a 2-D slowness map with 100 pixels (km) in both vertical and horizontal directions by re-sampling the original velocity model. For this model, We design two experiments with different receiver distribution. \emph{Case 1}) The number of receivers and straight rays on this map is the same as those on the smooth-discontinuous model, as are their locations (Fig.~\ref{marmousi_local}b). \emph{Case 2}) There are 100 receivers regularly distributed on on the slowness map (Fig.~\ref{marmousi_local}c). Additionally, we also test the traveltime with noise levels of $\sigma=0.02$ and $\sigma=0.05$, respectively. For the two cases, the number of atoms is fixed at 150 and other hyperparameters for LSQR, dictionary learning, and the proposed method are the same as those used in previous tests on the smooth-discontinuous model.

\begin{figure*}[]
\centering
\includegraphics[width=6in]{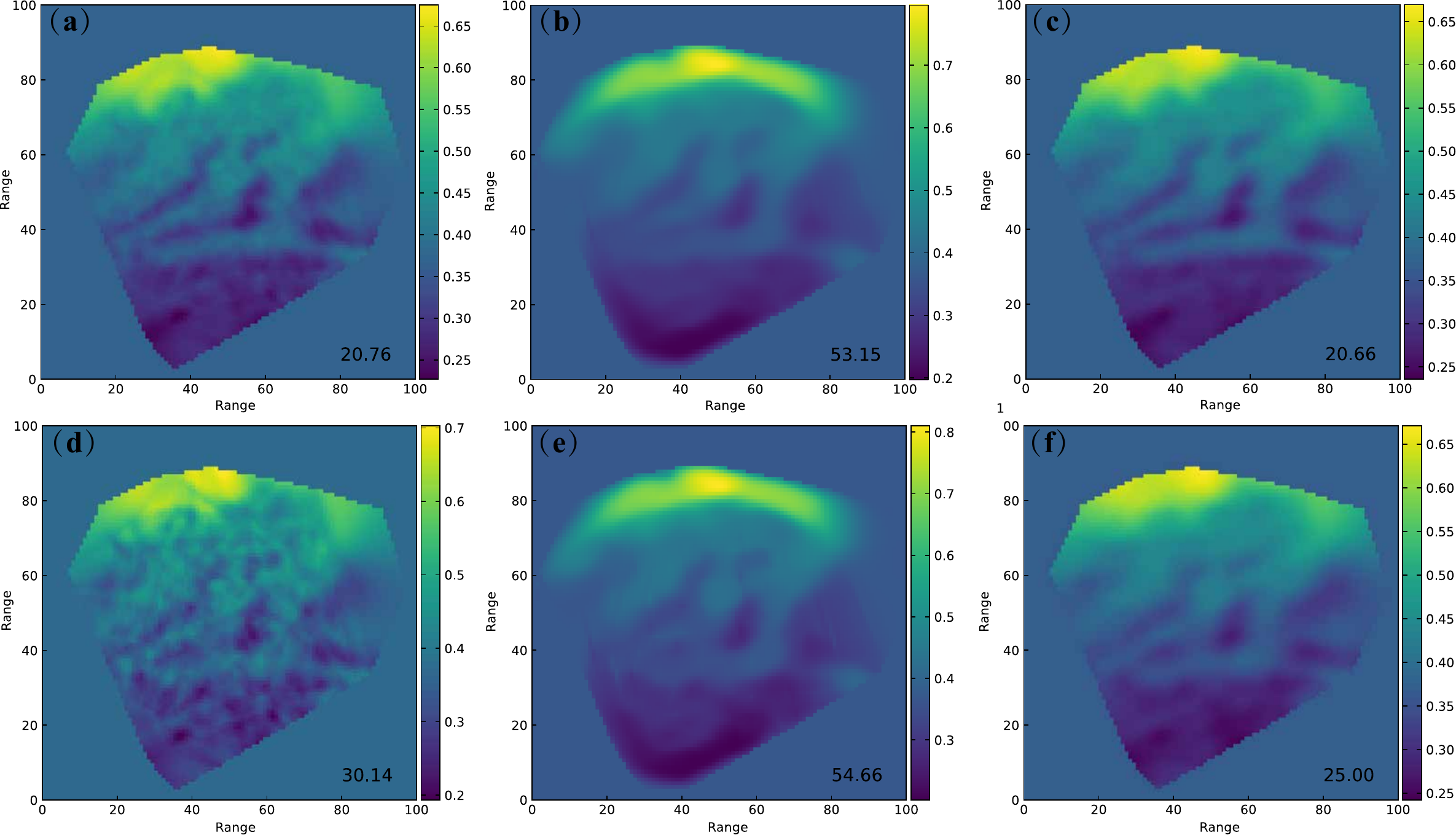}
\caption{Inverted slowness maps for the traveltime sampled by Fig.~\ref{marmousi_local}(b). (a)-(c) Slowness maps inverted by LSQR, dictionary learning, and the proposed method ($\sigma=0.02$). (d)-(f) Slowness maps inverted by LSQR, dictionary learning, and the proposed method ($\sigma=0.05$). RMSE values are printed on these slowness maps.}
\label{marmousi_inversion1}
\end{figure*}

\begin{figure*}[]
\centering
\includegraphics[width=5in]{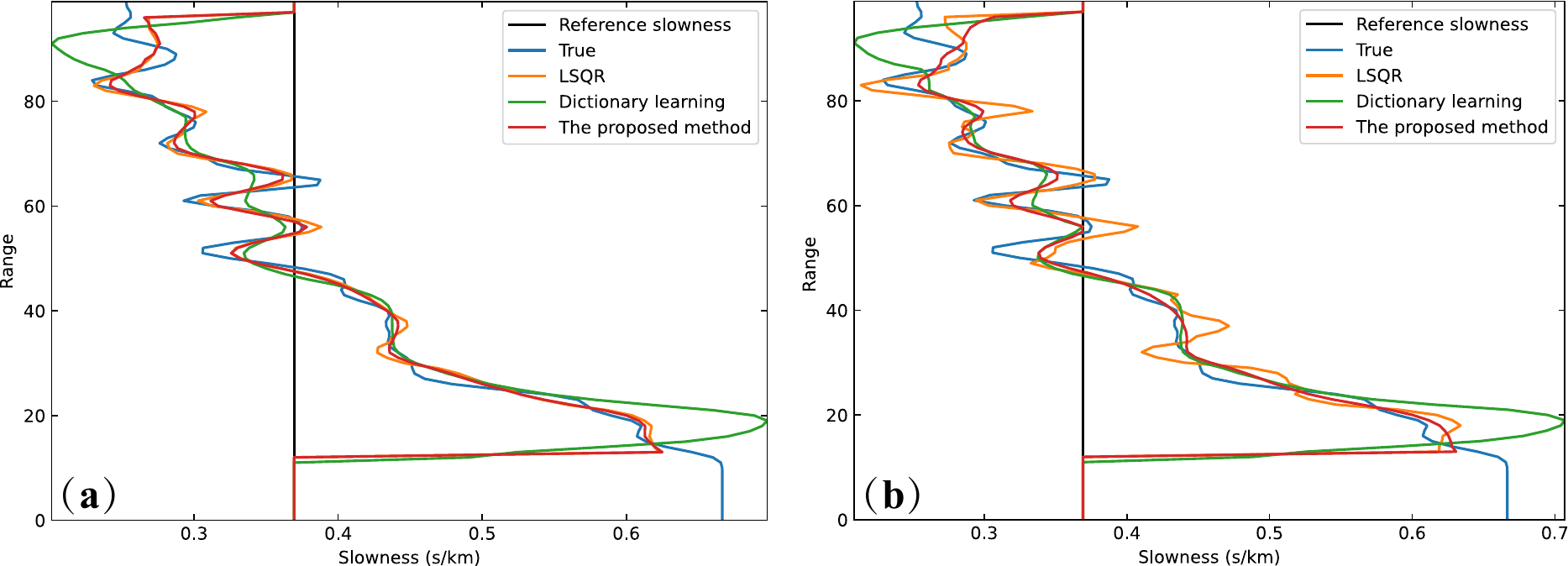}
\caption{Slowness profiles from Fig.~\ref{marmousi_inversion1}. (a) Comparison of 1-D slowness ($\sigma=0.02$). The RMSE values for LSQR, dictionary learning, and the proposed method are 80.72, 17.39, and 17.14, respectively. (b) Comparison of 1-D slowness ($\sigma=0.05$). The RMSE values for LSQR, dictionary learning, and the proposed method are 79.58, 36.39, and 29.13, respectively.}
\label{marmousi_slice1}
\end{figure*}

As shown in Fig.~\ref{marmousi_inversion1}a-c, the proposed method achieves higher resolution and lower RMSE than other two compared algorithms. Although the difference in RMSE between LSQR and the proposed method is slight, the slowness map inverted by our method is smoother than that of LSQR, and many details in the result inverted by dictionary learning are over-smoothed. As the noise level increases, our method still obtains higher resolution than these compared algorithms, while the resolution of results inverted by LSQR and dictionary learning decreases dramatically (Fig.~\ref{marmousi_inversion1}d-f). The difference in RMSE between our method and the compared algorithms increases significantly. The comparisons of 1-D slowness (Fig.~\ref{marmousi_slice1}a and Fig.~\ref{marmousi_slice1}b) further demonstrate the effectiveness and robustness of our method against noise impact.

\begin{figure*}[]
\centering
\includegraphics[width=6in]{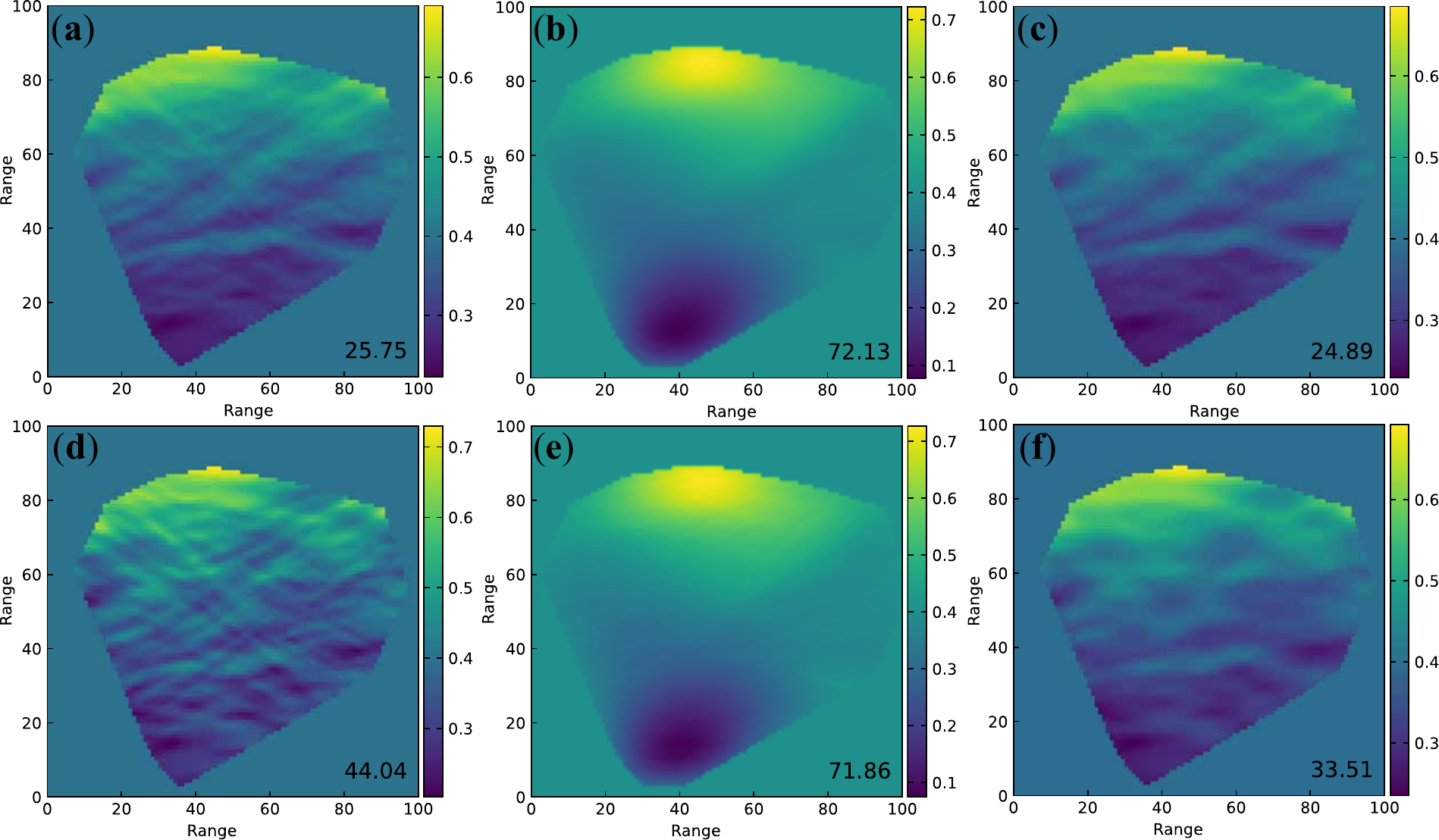}
\caption{Inverted slowness maps for the traveltime sampled by Fig.~\ref{marmousi_local}(b). (a)-(c) Slowness maps inverted by LSQR, dictionary learning, and the proposed method ($\sigma=0.02$). (d)-(f) Slowness maps inverted by LSQR, dictionary learning, and the proposed method ($\sigma=0.05$). RMSE values are printed on these slowness maps.}
\label{marmousi_inversion2}
\end{figure*}

\begin{figure*}[]
\centering
\includegraphics[width=5in]{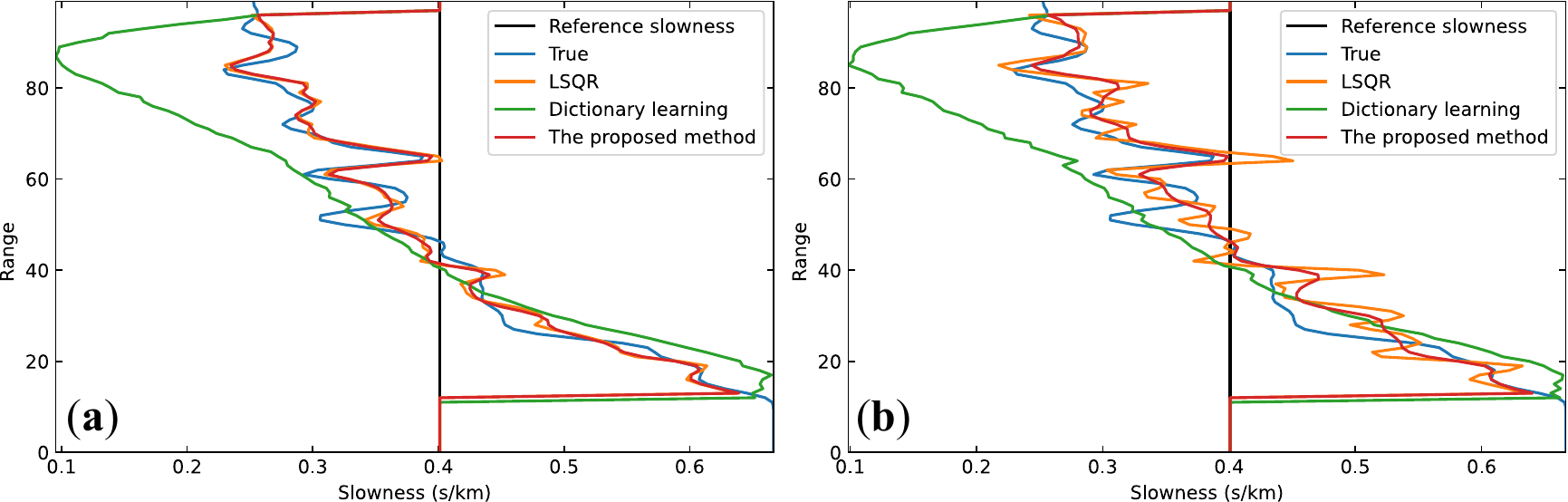}
\caption{Slowness profiles from Fig.~\ref{marmousi_inversion2}. (a) Comparison of 1-D slowness ($\sigma=0.02$). The RMSE values for LSQR, dictionary learning, and the proposed method are 13.82, 37.01, and 13.13, respectively. (b) Comparison of 1-D slowness ($\sigma=0.05$). The RMSE values for LSQR, dictionary learning, and the proposed method are 23.36, 37.56, and 18.63, respectively.}
\label{marmousi_slice2}
\end{figure*}

For case 2, one also can observe that the inverted results by the proposed method show higher resolution than other two compared algorithms (Fig.~\ref{marmousi_inversion2}). Specially, the results inverted by the proposed method are smoother than that of LSQR and preserve more details that that of dictionary learning. Moreover, as the noise level increasing, the quality of LSQR decreased rapidly, and dictionary learning only obtains the large-scale trend, while the proposed method exhibits good generalization for receiver distribution and robustness for different level of random noise (Fig.~\ref{marmousi_inversion2} and (Fig.~\ref{marmousi_slice2})).      

\subsection{Field traveltime}
\begin{figure*}[]
\centering
\includegraphics[width=5.5in]{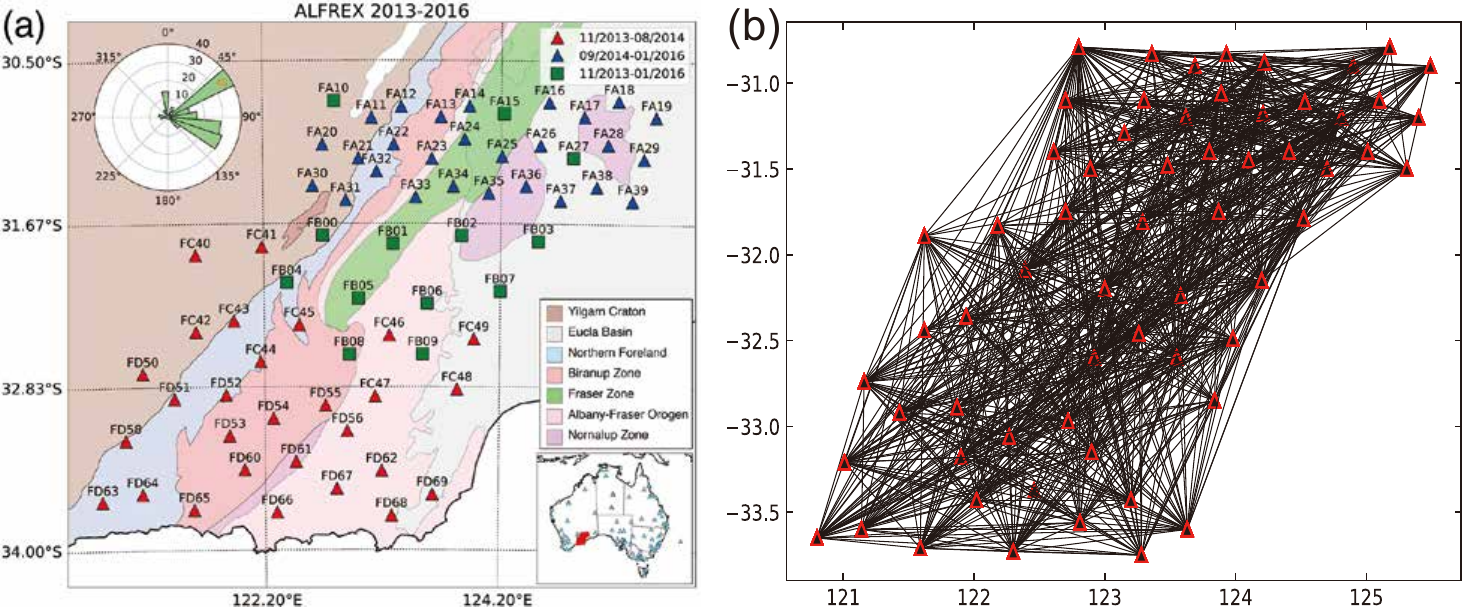}
\caption{(a) Spatiotemporal distribution of ALFREX seismic networks superimposed on regional geological maps of southern Australia. The crustal domains are colored to show the complex regional tectonic structures. The rose diagram shows the azimuthal distribution of the virtual source stations used in the empirical Green's function retrieval in the respective test cases. The radial axis is clipped for a better illustration, and the number of stations in the dominating direction is labeled on the bar, which is contributed from two dense arrays (Alice Springs and Warramunga arrays) in central Australia. In the inset map, the locations of permanent seismic stations acting as virtual sources are marked with the cyan triangles, and the ALFREX networks are highlighted in red; (b) The raypath coverages (875 rays) at 5 s of ambient noise fields cross-correlation functions from ALFREX. Only raypaths with robust traveltime measurements and distance greater than three times of Rayleigh wave wavelength are preserved \cite{chen_empirical_2020}.}
\label{ALDREX_geo_rays}
\end{figure*}

We further examine the the effectiveness of the proposed method using the real trvaltime obtained by ambient noise cross-correlation. The ambient noise data were recorded by the ALFREX network that consists of two subarrays, each sampling a part of the Albany-Fraser orogen in southwestern Australia at a different time, as well as 13 semipermanent stations operating throughout the acquisition period (Fig.~\ref{ALDREX_geo_rays}(a)). The raypaths between station pairs were derived by the Empirical Green's function obtained from source-receiver interferometry by measuring the Rayleigh wave traveltimes at period of 5 s (Fig.~\ref{ALDREX_geo_rays}(b)). For more details about the field traveltime, one can refer to reference \cite{chen_empirical_2020}. 

The study area is parameterized into a regular grid of $40\times 50$ nodes. For LSQR inversion, we set the $\eta$ and $L$ (Eq.~\ref{eq11}) to 5 $\text{km}^2$ and 10 $\text{km}$ respectively. The initial velocity is established using a constant value derived from $A^{-1}\times t$. For dictionary learning inversion, we allocate a damping coefficient of $=10$, a patch size of $4\times4$, and $1000$ iterations for the LSQR inversion. The number of iteration $k$ for the dictionary learning inversion is fixed at 150. We define the spares level $T_d$ and $H_d$ in Algorithm~\ref{alg2}, and the number of atoms for ITKM matches those used in the Marmousi model experiments, along with the specifics related to patch selection, and the design and training of the NN.  

\begin{figure}[]
\centering
\includegraphics[width=3.5in]{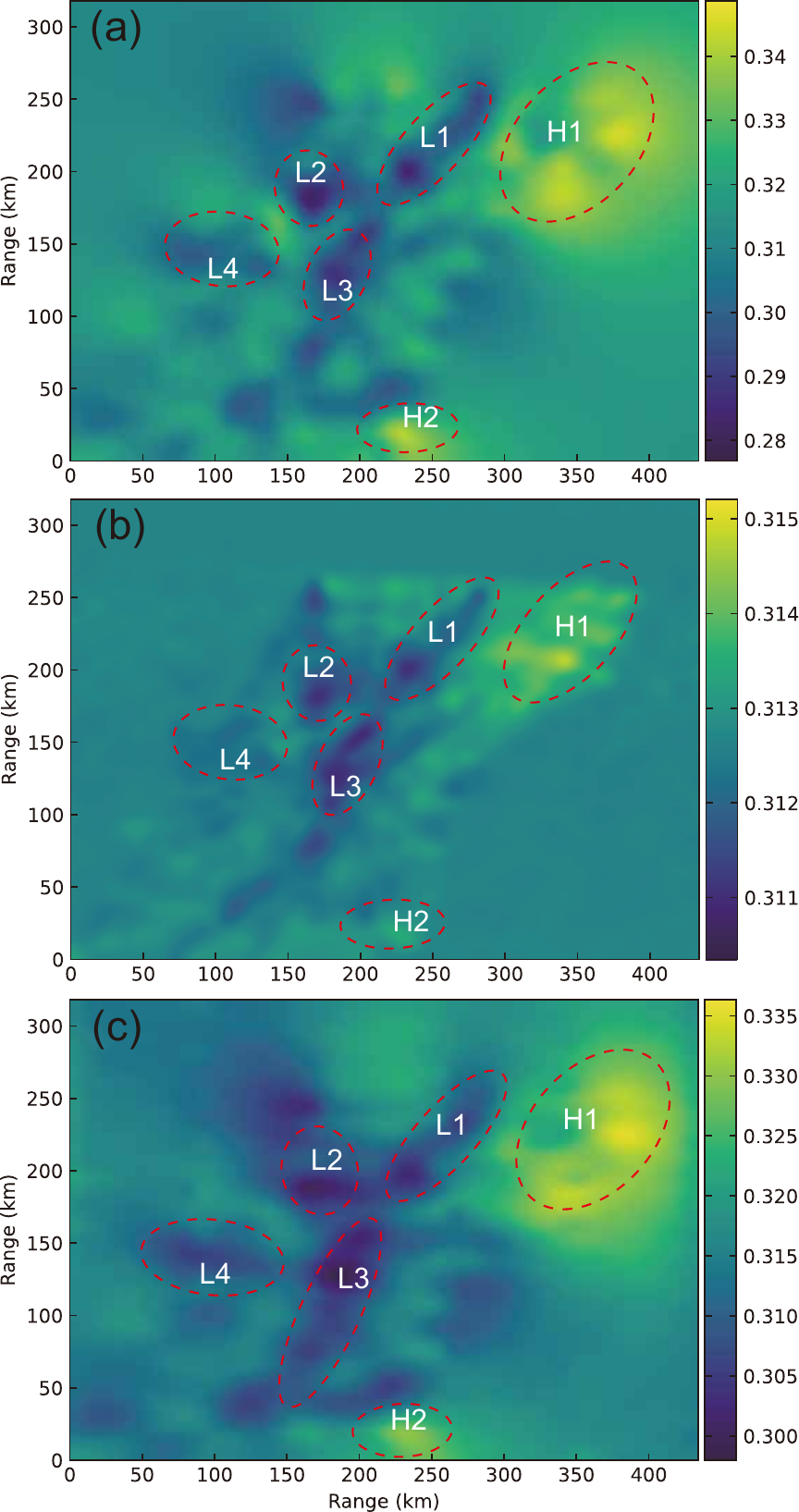}
\caption{Inverted slowness map of the traveltime obtained from ALFREX by (a) LSQR, (b) dictionary learning, and (c) the proposed method.}
\label{ALDREX_Inv}
\end{figure}

From the field traveltime tomographic results \ref{ALDREX_Inv}, we can obviously observe that the inversion result (Fig.~\ref{ALDREX_Inv}(a)) by the proposed method reveals four distinct low-slowness structures (L1-L4) and two high-slowness structures, consistent with the observations from the study by \citep{chen_empirical_2020} and \citep{sippl_crustal_2017}. Compared with the inversion results obtained by LSQR and dictionary learning (Fig.~\ref{ALDREX_Inv}(a) and (b)), the NE-SW striking high-velocity structure, marked by L1 and L3, is more clearly delineated by the proposed method. Also, the high-slowness structures H1 and H2 indicated by the proposed method are more pronounced than dictionary learning. Moreover, the proposed method provides a more distinct representation of the low-slowness structure L4 than LSQR and dictionary learning.

\section{Discussion}
\begin{figure*}[]
\centering
\includegraphics[width=6in]{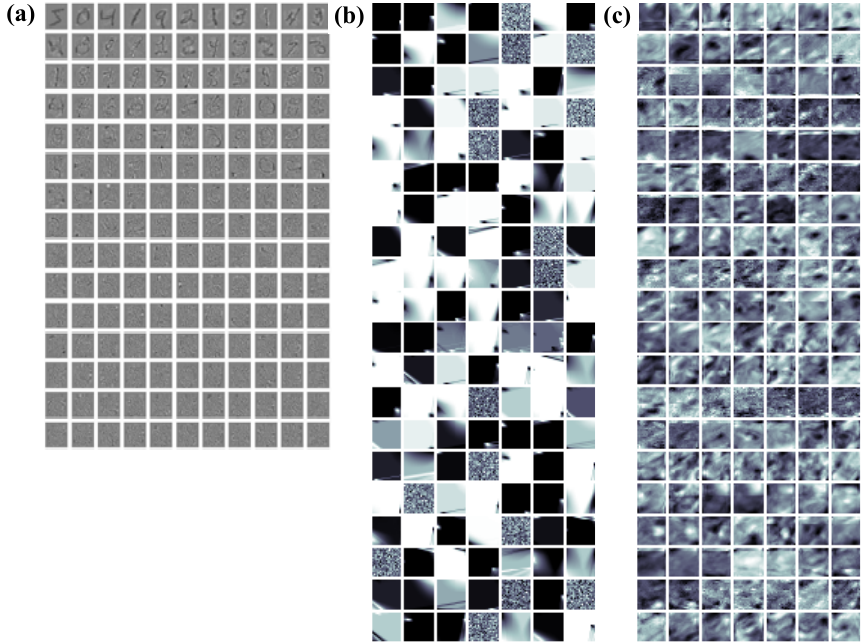}
\caption{(a) Dictionary trained by DDL with the MNIST dataset\cite{tariyal_deep_2016}. (c) Dictionary trained by dictionary learning on the Marmousi model test. (c) Dictionary optimized by the proposed method on the Marmousi model test. ($\sigma=0.05$, atoms=150).}
\label{dict}
\end{figure*}

End-to-end learning is the main manner in the current deep-learning-based tomography because it can be easily implemented and rapidly inferred. However, this learning steerage requires the labeled samples to train NNs. Real labels for field data inversion are usually missing or very expensive (e.g., well logging), limiting the application of deep learning in field data inversion. Therefore, how to develop a label-free learning inversion method is meaningful to field data inversion. In this paper, we propose to integrate dictionary learning and deep learning to enhance the resolution of LSQR estimation. In the proposed method, we train NNs by minimizing the MSE loss of traveltime using the initial dictionary and observed traveltime, which does not require to prepare the labeled samples through forwarding approaches or collecting logging data. 

On the other hand, our method can provide some guarantees for the reliability of the final inversion result. The NNs are used to optimize the dictionary instead of predicting VMs from observations, and the final slowness map is construct by summing the weighted reference slowness, the estimation by LSQR, and the optimized dictionary and corresponding code. Therefore, the role of NN can be considered to fine-tune the estimation by LSQR to obtain the high-resolution VMs, which may beneficial to mitigate uncertainty produced by the black-box nature of NNs in final VMs. In addition, the proposed method can provide the optimized dictionary after each epoch of NN training, reducing the computational cost. The computational cost of our method is low due to the few parameters of NNs and the small amount of training data (only including initial dictionary and observed traveltime), and it is dominated by the sparse level of the atoms in dictionary learning and of the code in sparse coding. Higher sparse levels result in lower computation costs.

The idea of PINN\cite{raissi_physics-informed_2019} is to embed the PDE into the loss function of NN training to reduce dependence on labeled samples. Deep Dictionary Learning (DDL) aims to learn multiple levels of dictionaries by combining deep learning and dictionary concepts \cite{tariyal_deep_2016}. Compared to the two algorithms, our method can not require the labeled samples to train NN. Although both DDL and our method combine deep learning and dictionary learning concepts, our method only needs to learn one dictionary, and it will be taken as the input for NN instead of being used as the "weight" or "filter" in conventional NNs in DDL. Furthermore, the NNs in our method are trained for the current traveltime tomography instead of training one model for many inversion tasks, which is beneficial to the reliability of inversion results. 

Dictionary learning is a powerful technique that focuses on refining patch-level or local feature of data. This process leads to the creation of dictionary atoms, which exhibit high sparsity due to the sparsity assumption enforced during the training phase. As a result, the learned atoms through the dictionary learning become highly sparse. In the Marmousi model test, for instance, the atoms mainly consists of curves and edges (Fig.~\ref{dict}b). Although the sparse dictionary can extract main information from a signal and suppress noise, it may sacrifice some details or weak signals and/or contain noise. 

In contrast to dictionary learning, our proposed method utilizes the NN to optimize the dictionary without making any assumptions about sparsity or over-completeness. This is crucial for ensuring good generalization. As shown in Fig.~\ref{dict}c, the features of the atoms achieved by our method are fundamentally different from those learned through dictionary learning and are much richer. Many atoms exhibit unique features that may even represent fundamental features not captured by curves and edges. We have observed the occurrence of new features in our previous study \cite{wang_learning_2022} and plan to further investigate this fascinating phenomenon in future studies. Furthermore, our method differs significantly from DDL since the optimized dictionary is the output of the NN in our method, while DDL replaces the “weight” or “filter” in conventional NNs with the dictionary. As a result, for both classification and clustering tasks, the atoms will hierarchically approximate the training labels (Fig.~\ref{dict}a).

\section{Conclusion}
In this article, we introduce a label-free tomographic method for seismic traveltime. Our approach integrates deep learning and dictionary learning to enhance the low-resolution VM inverted by the traditional tomographic algorithm-LSQR. We demonstrate the effectiveness of our method through numerical tests on both synthetic and field traveltime. Our method designs a shallow and simple NN and an optimized dictionary to train the NN without requiring labels. By minimizing the traveltime MSE loss using the initial dictionary and observed traveltime, the proposed method can provide the optimized dictionary after each epoch of NN training followed by reconstructing the high-resolution VM. The proposed tomography method exhibits potential for providing accurate initial VM for seismic imaging such as FWI and for deep learning-based geophysical inversion without real labels or training dataset.

\section*{Acknowledgment}
We thank Dr. Yunfeng Chen of Zhejiang University for providing the Rayleigh wave traveltimes obtained from ALFREX network in southwestern Australia.

\bibliographystyle{IEEEtran}
\bibliography{tomo}

\begin{thebibliography}{10}
\providecommand{\url}[1]{#1}
\csname url@samestyle\endcsname
\providecommand{\newblock}{\relax}
\providecommand{\bibinfo}[2]{#2}
\providecommand{\BIBentrySTDinterwordspacing}{\spaceskip=0pt\relax}
\providecommand{\BIBentryALTinterwordstretchfactor}{4}
\providecommand{\BIBentryALTinterwordspacing}{\spaceskip=\fontdimen2\font plus
\BIBentryALTinterwordstretchfactor\fontdimen3\font minus \fontdimen4\font\relax}
\providecommand{\BIBforeignlanguage}[2]{{%
\expandafter\ifx\csname l@#1\endcsname\relax
\typeout{** WARNING: IEEEtran.bst: No hyphenation pattern has been}%
\typeout{** loaded for the language `#1'. Using the pattern for}%
\typeout{** the default language instead.}%
\else
\language=\csname l@#1\endcsname
\fi
#2}}
\providecommand{\BIBdecl}{\relax}
\BIBdecl

\bibitem{mordret_ambient_2014}
\BIBentryALTinterwordspacing
A.~Mordret, M.~Landès, N.~M. Shapiro, S.~C. Singh, and P.~Roux, ``\BIBforeignlanguage{en}{Ambient noise surface wave tomography to determine the shallow shear velocity structure at {Valhall}: depth inversion with a {Neighbourhood} {Algorithm}},'' \emph{\BIBforeignlanguage{en}{Geophysical Journal International}}, vol. 198, no.~3, pp. 1514--1525, Sep. 2014. [Online]. Available: \url{http://academic.oup.com/gji/article/198/3/1514/587419/Ambient-noise-surface-wave-tomography-to-determine}
\BIBentrySTDinterwordspacing

\bibitem{gorbatov_signature_2000}
\BIBentryALTinterwordspacing
A.~Gorbatov, S.~Widiyantoro, Y.~Fukao, and E.~Gordeev, ``\BIBforeignlanguage{en}{Signature of remnant slabs in the {North} {Pacific} from {P}-wave tomography},'' \emph{\BIBforeignlanguage{en}{Geophysical Journal International}}, vol. 142, no.~1, pp. 27--36, Jul. 2000. [Online]. Available: \url{https://academic.oup.com/gji/article-lookup/doi/10.1046/j.1365-246x.2000.00122.x}
\BIBentrySTDinterwordspacing

\bibitem{meier_global_2007}
\BIBentryALTinterwordspacing
U.~Meier, A.~Curtis, and J.~Trampert, ``\BIBforeignlanguage{en}{Global crustal thickness from neural network inversion of surface wave data},'' \emph{\BIBforeignlanguage{en}{Geophysical Journal International}}, vol. 169, no.~2, pp. 706--722, May 2007. [Online]. Available: \url{https://academic.oup.com/gji/article-lookup/doi/10.1111/j.1365-246X.2007.03373.x}
\BIBentrySTDinterwordspacing

\bibitem{allmark_seismic_2018}
\BIBentryALTinterwordspacing
C.~Allmark, A.~Curtis, E.~Galetti, and S.~Ridder, ``\BIBforeignlanguage{en}{Seismic {Attenuation} {From} {Ambient} {Noise} {Across} the {North} {Sea} {Ekofisk} {Permanent} {Array}},'' \emph{\BIBforeignlanguage{en}{Journal of Geophysical Research: Solid Earth}}, vol. 123, no.~10, pp. 8691--8710, Oct. 2018. [Online]. Available: \url{https://onlinelibrary.wiley.com/doi/10.1029/2017JB015419}
\BIBentrySTDinterwordspacing

\bibitem{loris_nonlinear_2010}
\BIBentryALTinterwordspacing
I.~Loris, H.~Douma, G.~Nolet, I.~Daubechies, and C.~Regone, ``Nonlinear regularization techniques for seismic tomography,'' \emph{Journal of Computational Physics}, vol. 229, no.~3, pp. 890--905, Feb. 2010, arXiv:0808.3472 [physics]. [Online]. Available: \url{http://arxiv.org/abs/0808.3472}
\BIBentrySTDinterwordspacing

\bibitem{bianco_travel_2018}
M.~J. Bianco and P.~Gerstoft, ``Travel {Time} {Tomography} {With} {Adaptive} {Dictionaries},'' \emph{IEEE Transactions on Computational Imaging}, vol.~4, no.~4, pp. 499--511, 2018, conference Name: IEEE Transactions on Computational Imaging.

\bibitem{galetti_uncertainty_2015}
\BIBentryALTinterwordspacing
E.~Galetti, A.~Curtis, G.~A. Meles, and B.~Baptie, ``Uncertainty {Loops} in {Travel}-{Time} {Tomography} from {Nonlinear} {Wave} {Physics},'' \emph{Physical Review Letters}, vol. 114, no.~14, p. 148501, Apr. 2015, publisher: American Physical Society. [Online]. Available: \url{https://link.aps.org/doi/10.1103/PhysRevLett.114.148501}
\BIBentrySTDinterwordspacing

\bibitem{piana_agostinetti_local_2015}
\BIBentryALTinterwordspacing
N.~Piana~Agostinetti, G.~Giacomuzzi, and A.~Malinverno, ``Local three-dimensional earthquake tomography by trans-dimensional {Monte} {Carlo} sampling,'' \emph{Geophysical Journal International}, vol. 201, no.~3, pp. 1598--1617, Jun. 2015. [Online]. Available: \url{https://doi.org/10.1093/gji/ggv084}
\BIBentrySTDinterwordspacing

\bibitem{zhao_bayesian_2022}
\BIBentryALTinterwordspacing
X.~Zhao, A.~Curtis, and X.~Zhang, ``Bayesian seismic tomography using normalizing flows,'' \emph{Geophysical Journal International}, vol. 228, no.~1, pp. 213--239, Jan. 2022. [Online]. Available: \url{https://doi.org/10.1093/gji/ggab298}
\BIBentrySTDinterwordspacing

\bibitem{mousavi_deep-learning_2022}
\BIBentryALTinterwordspacing
S.~M. Mousavi and G.~C. Beroza, ``\BIBforeignlanguage{en}{Deep-learning seismology},'' \emph{\BIBforeignlanguage{en}{Science}}, vol. 377, no. 6607, p. eabm4470, Aug. 2022. [Online]. Available: \url{https://www.science.org/doi/10.1126/science.abm4470}
\BIBentrySTDinterwordspacing

\bibitem{moya_inversion_2010}
\BIBentryALTinterwordspacing
A.~Moya and K.~Irikura, ``\BIBforeignlanguage{en}{Inversion of a velocity model using artificial neural networks},'' \emph{\BIBforeignlanguage{en}{Computers \& Geosciences}}, vol.~36, no.~12, pp. 1474--1483, Dec. 2010. [Online]. Available: \url{https://www.sciencedirect.com/science/article/pii/S0098300410000221}
\BIBentrySTDinterwordspacing

\bibitem{araya-polo_deep-learning_2018}
\BIBentryALTinterwordspacing
M.~Araya-Polo, J.~Jennings, A.~Adler, and T.~Dahlke, ``\BIBforeignlanguage{en}{Deep-learning tomography},'' \emph{\BIBforeignlanguage{en}{The Leading Edge}}, vol.~37, no.~1, pp. 58--66, Jan. 2018. [Online]. Available: \url{https://library.seg.org/doi/10.1190/tle37010058.1}
\BIBentrySTDinterwordspacing

\bibitem{geng_deep_2022}
\BIBentryALTinterwordspacing
Z.~Geng, Z.~Zhao, Y.~Shi, X.~Wu, S.~Fomel, and M.~Sen, ``Deep learning for velocity model building with common-image gather volumes,'' \emph{Geophysical Journal International}, vol. 228, no.~2, pp. 1054--1070, Feb. 2022. [Online]. Available: \url{https://doi.org/10.1093/gji/ggab385}
\BIBentrySTDinterwordspacing

\bibitem{fabien-ouellet_seismic_2020}
\BIBentryALTinterwordspacing
G.~Fabien-Ouellet and R.~Sarkar, ``\BIBforeignlanguage{en}{Seismic velocity estimation: {A} deep recurrent neural-network approach},'' \emph{\BIBforeignlanguage{en}{GEOPHYSICS}}, vol.~85, no.~1, pp. U21--U29, Jan. 2020. [Online]. Available: \url{https://library.seg.org/doi/10.1190/geo2018-0786.1}
\BIBentrySTDinterwordspacing

\bibitem{goodfellow_deep_2016}
I.~Goodfellow, Y.~Bengio, and A.~Courville, \emph{Deep {Learning}}.\hskip 1em plus 0.5em minus 0.4em\relax MIT Press, 2016.

\bibitem{cai_semisupervised_2022}
\BIBentryALTinterwordspacing
A.~Cai, H.~Qiu, and F.~Niu, ``\BIBforeignlanguage{en}{Semi‐{Supervised} {Surface} {Wave} {Tomography} {With} {Wasserstein} {Cycle}‐{Consistent} {GAN}: {Method} and {Application} to {Southern} {California} {Plate} {Boundary} {Region}},'' \emph{\BIBforeignlanguage{en}{Journal of Geophysical Research: Solid Earth}}, vol. 127, no.~3, Mar. 2022. [Online]. Available: \url{https://onlinelibrary.wiley.com/doi/10.1029/2021JB023598}
\BIBentrySTDinterwordspacing

\bibitem{devilee_efficient_1999}
\BIBentryALTinterwordspacing
R.~J.~R. Devilee, A.~Curtis, and K.~Roy-Chowdhury, ``\BIBforeignlanguage{en}{An efficient, probabilistic neural network approach to solving inverse problems: {Inverting} surface wave velocities for {Eurasian} crustal thickness},'' \emph{\BIBforeignlanguage{en}{Journal of Geophysical Research: Solid Earth}}, vol. 104, no. B12, pp. 28\,841--28\,857, Dec. 1999. [Online]. Available: \url{http://doi.wiley.com/10.1029/1999JB900273}
\BIBentrySTDinterwordspacing

\bibitem{earp_probabilistic_2020}
\BIBentryALTinterwordspacing
S.~Earp and A.~Curtis, ``\BIBforeignlanguage{en}{Probabilistic neural network-based {2D} travel-time tomography},'' \emph{\BIBforeignlanguage{en}{Neural Computing and Applications}}, vol.~32, no.~22, pp. 17\,077--17\,095, Nov. 2020. [Online]. Available: \url{https://doi.org/10.1007/s00521-020-04921-8}
\BIBentrySTDinterwordspacing

\bibitem{raissi_physics-informed_2019}
\BIBentryALTinterwordspacing
M.~Raissi, P.~Perdikaris, and G.~E. Karniadakis, ``\BIBforeignlanguage{en}{Physics-informed neural networks: {A} deep learning framework for solving forward and inverse problems involving nonlinear partial differential equations},'' \emph{\BIBforeignlanguage{en}{Journal of Computational Physics}}, vol. 378, pp. 686--707, Feb. 2019. [Online]. Available: \url{http://www.sciencedirect.com/science/article/pii/S0021999118307125}
\BIBentrySTDinterwordspacing

\bibitem{song_solving_2021}
\BIBentryALTinterwordspacing
C.~Song, T.~Alkhalifah, and U.~B. Waheed, ``Solving the frequency-domain acoustic {VTI} wave equation using physics-informed neural networks,'' \emph{Geophysical Journal International}, vol. 225, no.~2, pp. 846--859, Apr. 2021. [Online]. Available: \url{https://doi.org/10.1093/gji/ggab010}
\BIBentrySTDinterwordspacing

\bibitem{song_simulating_2022}
\BIBentryALTinterwordspacing
C.~Song and Y.~Wang, ``\BIBforeignlanguage{en}{Simulating seismic multifrequency wavefields with the {Fourier} feature physics-informed neural network},'' \emph{\BIBforeignlanguage{en}{Geophysical Journal International}}, vol. 232, no.~3, pp. 1503--1514, Nov. 2022. [Online]. Available: \url{https://academic.oup.com/gji/article/232/3/1503/6758508}
\BIBentrySTDinterwordspacing

\bibitem{waheed_pinneik_2021}
\BIBentryALTinterwordspacing
U.~b. Waheed, E.~Haghighat, T.~Alkhalifah, C.~Song, and Q.~Hao, ``\BIBforeignlanguage{en}{{PINNeik}: {Eikonal} solution using physics-informed neural networks},'' \emph{\BIBforeignlanguage{en}{Computers \& Geosciences}}, vol. 155, p. 104833, Oct. 2021. [Online]. Available: \url{https://www.sciencedirect.com/science/article/pii/S009830042100131X}
\BIBentrySTDinterwordspacing

\bibitem{taufik_upwind_2022}
M.~H. Taufik, U.~b. Waheed, and T.~A. Alkhalifah, ``Upwind, {No} {More}: {Flexible} {Traveltime} {Solutions} {Using} {Physics}-{Informed} {Neural} {Networks},'' \emph{IEEE Transactions on Geoscience and Remote Sensing}, vol.~60, pp. 1--12, 2022, conference Name: IEEE Transactions on Geoscience and Remote Sensing.

\bibitem{gou_bayesian_2022}
\BIBentryALTinterwordspacing
R.~Gou, Y.~Zhang, X.~Zhu, and J.~Gao, ``Bayesian {Physics}-{Informed} {Neural} {Networks} for the {Subsurface} {Tomography} based on the {Eikonal} {Equation},'' Nov. 2022, arXiv:2203.12351 [physics]. [Online]. Available: \url{http://arxiv.org/abs/2203.12351}
\BIBentrySTDinterwordspacing

\bibitem{grubas_neural_2023}
\BIBentryALTinterwordspacing
S.~Grubas, A.~Duchkov, and G.~Loginov, ``\BIBforeignlanguage{en}{Neural {Eikonal} solver: {Improving} accuracy of physics-informed neural networks for solving eikonal equation in case of caustics},'' \emph{\BIBforeignlanguage{en}{Journal of Computational Physics}}, vol. 474, p. 111789, Feb. 2023. [Online]. Available: \url{https://www.sciencedirect.com/science/article/pii/S002199912200852X}
\BIBentrySTDinterwordspacing

\bibitem{chen_eikonal_2022}
\BIBentryALTinterwordspacing
Y.~Chen, S.~A.~L. de~Ridder, S.~Rost, Z.~Guo, X.~Wu, and Y.~Chen, ``\BIBforeignlanguage{en}{Eikonal {Tomography} {With} {Physics}-{Informed} {Neural} {Networks}: {Rayleigh} {Wave} {Phase} {Velocity} in the {Northeastern} {Margin} of the {Tibetan} {Plateau}},'' \emph{\BIBforeignlanguage{en}{Geophysical Research Letters}}, vol.~49, no.~21, p. e2022GL099053, 2022, \_eprint: https://onlinelibrary.wiley.com/doi/pdf/10.1029/2022GL099053. [Online]. Available: \url{https://onlinelibrary.wiley.com/doi/abs/10.1029/2022GL099053}
\BIBentrySTDinterwordspacing

\bibitem{tariyal_deep_2016}
S.~Tariyal, A.~Majumdar, R.~Singh, and M.~Vatsa, ``Deep {Dictionary} {Learning},'' \emph{IEEE Access}, vol.~4, pp. 10\,096--10\,109, 2016, conference Name: IEEE Access.

\bibitem{mahdizadehaghdam_deep_2019}
\BIBentryALTinterwordspacing
S.~Mahdizadehaghdam, A.~Panahi, H.~Krim, and L.~Dai, ``Deep {Dictionary} {Learning}: {A} {PARametric} {NETwork} {Approach},'' \emph{IEEE Transactions on Image Processing}, vol.~28, no.~10, pp. 4790--4802, Oct. 2019. [Online]. Available: \url{https://ieeexplore.ieee.org/document/8708973/}
\BIBentrySTDinterwordspacing

\bibitem{tang_when_2021}
\BIBentryALTinterwordspacing
H.~Tang, H.~Liu, W.~Xiao, and N.~Sebe, ``When {Dictionary} {Learning} {Meets} {Deep} {Learning}: {Deep} {Dictionary} {Learning} and {Coding} {Network} for {Image} {Recognition} {With} {Limited} {Data},'' \emph{IEEE Transactions on Neural Networks and Learning Systems}, vol.~32, no.~5, pp. 2129--2141, May 2021. [Online]. Available: \url{https://ieeexplore.ieee.org/document/9112646/}
\BIBentrySTDinterwordspacing

\bibitem{nicolson_rayleigh_2014}
\BIBentryALTinterwordspacing
H.~Nicolson, A.~Curtis, and B.~Baptie, ``\BIBforeignlanguage{en}{Rayleigh wave tomography of the {British} {Isles} from ambient seismic noise},'' \emph{\BIBforeignlanguage{en}{Geophysical Journal International}}, vol. 198, no.~2, pp. 637--655, Aug. 2014. [Online]. Available: \url{http://academic.oup.com/gji/article/198/2/637/594945/Rayleigh-wave-tomography-of-the-British-Isles-from}
\BIBentrySTDinterwordspacing

\bibitem{galetti_transdimensional_2017}
\BIBentryALTinterwordspacing
E.~Galetti, A.~Curtis, B.~Baptie, D.~Jenkins, and H.~Nicolson, ``\BIBforeignlanguage{en}{Transdimensional {Love}-wave tomography of the {British} {Isles} and shear-velocity structure of the {East} {Irish} {Sea} {Basin} from ambient-noise interferometry},'' \emph{\BIBforeignlanguage{en}{Geophysical Journal International}}, vol. 208, no.~1, pp. 36--58, Jan. 2017. [Online]. Available: \url{https://academic.oup.com/gji/article-lookup/doi/10.1093/gji/ggw286}
\BIBentrySTDinterwordspacing

\bibitem{snieder_equivalence_2006}
\BIBentryALTinterwordspacing
R.~Snieder, J.~Sheiman, and R.~Calvert, ``\BIBforeignlanguage{en}{Equivalence of the virtual-source method and wave-field deconvolution in seismic interferometry},'' \emph{\BIBforeignlanguage{en}{Physical Review E}}, vol.~73, no.~6, p. 066620, Jun. 2006. [Online]. Available: \url{https://link.aps.org/doi/10.1103/PhysRevE.73.066620}
\BIBentrySTDinterwordspacing

\bibitem{rodgers_inverse_2008}
C.~D. Rodgers, \emph{\BIBforeignlanguage{eng}{Inverse methods for atmospheric sounding: theory and practice}}, repr~ed., ser. Series on atmospheric, oceanic and planetary physics.\hskip 1em plus 0.5em minus 0.4em\relax Singapore: World Scientific, 2008, no.~2.

\bibitem{tarantola_inverse_2005}
A.~Tarantola, \emph{\BIBforeignlanguage{English}{Inverse problem theory and methods for model paramenter estimation}}.\hskip 1em plus 0.5em minus 0.4em\relax Philadelphia, PA: Society for Industrial and Applied Mathematics, 2005, oCLC: 946579460.

\bibitem{schnass_local_2015}
\BIBentryALTinterwordspacing
K.~Schnass, ``Local {Identification} of {Overcomplete} {Dictionaries},'' Apr. 2015, arXiv:1401.6354 [cs, math, stat]. [Online]. Available: \url{http://arxiv.org/abs/1401.6354}
\BIBentrySTDinterwordspacing

\bibitem{pati_orthogonal_1993}
\BIBentryALTinterwordspacing
Y.~Pati, R.~Rezaiifar, and P.~Krishnaprasad, ``Orthogonal matching pursuit: recursive function approximation with applications to wavelet decomposition,'' in \emph{Proceedings of 27th {Asilomar} {Conference} on {Signals}, {Systems} and {Computers}}.\hskip 1em plus 0.5em minus 0.4em\relax Pacific Grove, CA, USA: IEEE Comput. Soc. Press, 1993, pp. 40--44. [Online]. Available: \url{http://ieeexplore.ieee.org/document/342465/}
\BIBentrySTDinterwordspacing

\bibitem{zhou_high-resolution_2022}
\BIBentryALTinterwordspacing
Z.~Zhou, M.~Bianco, P.~Gerstoft, and K.~Olsen, ``\BIBforeignlanguage{en}{High-{Resolution} {Imaging} of {Complex} {Shallow} {Fault} {Zones} {Along} the {July} 2019 {Ridgecrest} {Ruptures}},'' \emph{\BIBforeignlanguage{en}{Geophysical Research Letters}}, vol.~49, no.~1, p. e2021GL095024, 2022, \_eprint: https://onlinelibrary.wiley.com/doi/pdf/10.1029/2021GL095024. [Online]. Available: \url{https://onlinelibrary.wiley.com/doi/abs/10.1029/2021GL095024}
\BIBentrySTDinterwordspacing

\bibitem{mairal_sparse_2014}
\BIBentryALTinterwordspacing
J.~Mairal, ``\BIBforeignlanguage{en}{Sparse {Modeling} for {Image} and {Vision} {Processing}},'' \emph{\BIBforeignlanguage{en}{Foundations and Trends® in Computer Graphics and Vision}}, vol.~8, no. 2-3, pp. 85--283, 2014. [Online]. Available: \url{http://www.nowpublishers.com/articles/foundations-and-trends-in-computer-graphics-and-vision/CGV-058}
\BIBentrySTDinterwordspacing

\bibitem{chen_empirical_2020}
\BIBentryALTinterwordspacing
Y.~Chen and E.~Saygin, ``\BIBforeignlanguage{en}{Empirical {Green}'s {Function} {Retrieval} {Using} {Ambient} {Noise} {Source}‐{Receiver} {Interferometry}},'' \emph{\BIBforeignlanguage{en}{Journal of Geophysical Research: Solid Earth}}, vol. 125, no.~2, Feb. 2020. [Online]. Available: \url{https://onlinelibrary.wiley.com/doi/abs/10.1029/2019JB018261}
\BIBentrySTDinterwordspacing

\bibitem{sippl_crustal_2017}
\BIBentryALTinterwordspacing
C.~Sippl, B.~Kennett, H.~Tkalčić, K.~Gessner, and C.~Spaggiari, ``Crustal surface wave velocity structure of the east {Albany}-{Fraser} {Orogen}, {Western} {Australia}, from ambient noise recordings,'' \emph{Geophysical Journal International}, vol. 210, no.~3, pp. 1641--1651, Sep. 2017. [Online]. Available: \url{https://doi.org/10.1093/gji/ggx264}
\BIBentrySTDinterwordspacing

\bibitem{wang_learning_2022}
F.~Wang, B.~Yang, Y.~Wang, and M.~Wang, ``Learning {From} {Noisy} {Data}: {An} {Unsupervised} {Random} {Denoising} {Method} for {Seismic} {Data} {Using} {Model}-{Based} {Deep} {Learning},'' \emph{IEEE Transactions on Geoscience and Remote Sensing}, vol.~60, pp. 1--14, 2022, conference Name: IEEE Transactions on Geoscience and Remote Sensing.

\end{thebibliography}
\end{document}